\documentclass[final,3p]{elsarticle}
\usepackage{amsmath}
\usepackage{amssymb}
\usepackage{amsthm}

\usepackage[pdfpagemode=UseNone,pdfstartview=FitH]{hyperref}

\usepackage{lineno}

\journal{Applied Mathematics and Computation}
\usepackage{showkeys}
\usepackage{array}
\usepackage{color}
\usepackage{rotating}
\usepackage{tabularx}

\newcommand{\PreserveBackslash}[1]{\let\temp=\\#1\let\\=\temp}
\let\PBS=\PreserveBackslash
\newcommand{\cecol}{\PBS\centering\hspace{0pt}}  
\newcommand{\rrcol}{\PBS\raggedright\hspace{0pt}}
\newcommand{\rlcol}{\PBS\raggedleft\hspace{0pt}} 
\newcolumntype{R}[1]{>{\rrcol}p{#1}}
\newcolumntype{L}[1]{>{\rlcol}p{#1}}
\newcolumntype{C}[1]{>{\cecol}p{#1}}

\usepackage{stmaryrd}

\begin{document}

\begin{frontmatter}



  \title{Immersed boundary simulations of fluid shear-induced
    deformation of a cantilever beam}
  %


\author[amity]{Sudeshna Ghosh\corref{cor1}}
\ead{sudeshnagh108@gmail.com}

\address[amity]{Department of Mathematics, Amity School of Applied Sciences, Amity University Haryana, Gurugram}




\cortext[cor1]{Corresponding author}



%
%

\begin{abstract}
  We derive a mathematical model and the corresponding computational
  scheme to study deflection of a two-dimensional elastic cantilever
  beam immersed in a channel, where one end of the beam is fixed to the
  channel wall. The immersed boundary method has been employed to
  simulate numerically the fluid-structure interaction problem. We
  investigate how variations in physical and numerical parameters change
  the effective material properties of the elastic beam and compare the
  results qualitatively with linear beam theory. We also pay careful
  attention to ``corner effects'' -- irregularities in beam shape near
  the free and fixed ends -- and show how this can be remedied by
  smoothing out the corners with a ``fillet'' or rounded shape. Finally,
  we extend the immersed boundary formulation to include porosity in the
  beam and investigate the effect that the resultant porous flow has on
  beam deflection.
\end{abstract}


\end{frontmatter}


\newcommand{\en}[1]{(\ref{eq:#1})}
\newcommand{\leavethisout}[1]{}

\newcommand{\etal}{{\itshape et al.}}
\newcommand{\dt}{\Delta t}
\newcommand{\dx}{h_x}
\newcommand{\dy}{h_y}
\newcommand{\Reynolds}{\text{\emph{Re}}}
\newcommand{\units}[1]{\mbox{$\mathrm{#1}$}}
\newcommand{\bunits}[1]{[\mbox{$\mathrm{#1}$}]}
\newcommand{\starone}{(1)}
\newcommand{\startwo}{(2)}
\newcommand{\starthree}{(3)}
\newcommand{\fsub}[2]{#1_{\mbox{}\!\text{\scriptsize\emph{#2}}}}

\newcommand{\myvec}[1]{\vec{#1}}
\renewcommand{\myvec}[1]{\boldsymbol{#1}}
\newcommand{\vd}{\myvec{d}}
\newcommand{\vX}{\myvec{X}}
\newcommand{\vU}{\myvec{U}}
\newcommand{\vF}{\myvec{F}}
\newcommand{\vu}{\myvec{u}}
\newcommand{\vx}{\myvec{x}}
\newcommand{\vf}{\myvec{f}}
\newcommand{\vq}{\myvec{q}}
\newcommand{\vn}{\myvec{t}}
\newcommand{\vt}{\myvec{n}}
\newcommand{\Fwall}{\vF^{w}}
\newcommand{\Fwallb}{\vF^{w1}}
\newcommand{\Fwallt}{\vF^{w2}}
\newcommand{\Fbeam}{\vF^{b}}
\newcommand{\Fattach}{\vF^{a}}
\newcommand{\Xwall}{\vX^{w}}
\newcommand{\Xwallb}{\vX^{w1}}
\newcommand{\Xwallt}{\vX^{w2}}
\newcommand{\Hbeam}{H_b}
\newcommand{\Wbeam}{W_b}
\newcommand{\dsep}{h}
\newcommand{\deflect}{d}
\newcommand{\utop}{u_{\text{\emph{top}}}}
\newcommand{\EIeff}{EI_{\text{\emph{eff}}}}
\newcommand{\jump}[1]{\llbracket #1 \rrbracket}
\newcommand{\hwall}{h_{w}}

\newtheorem{note}{Note}


\section{Introduction}
\label{sec:intro}

In this paper, we derive a mathematical model and the corresponding
computational scheme to study the deflection of a two-dimensional
deformable elastic cantilever beam in response to a surrounding fluid
flow.  We will study cantilever beam immersed in a channel flow, where
one end of the beam is fixed to the channel wall. We will not impose a
given shear flow but rather drive the motion via two walls: one fixed
and one moving with constant velocity. In the absence of a beam or other
channel obstruction, the flow would develop into a simple linear shear
flow (also called Couette flow).  We also study the effect of
introducing a small porosity into the beam structure, and compare the
flow over porous and solid (non-porous) beams.  The choice of such a
porous cantilever is motivated by our interest in modelling biofilm
structures in the near future \cite{alpkvist-klapper-2007}. A biofilm is a collection of
microorganisms (immersed in fluid) that adhere to each other and often
also on a nearby surface. Fluid-structure interaction (FSI) plays a key
role in several phases of the biofilm life cycle, one being in the
deformation of biofilm columns due to fluid forces. We extract from this
scenario an idealized 2D model problem in which a rectangular cantilever
beam deforms in response to a shear flow. The method we develop may also
be useful for simulating other systems arising in applications from
biology (e.g., glycocalyx, cilia) and engineering (e.g., MEMS,
micropillars).

The study of a cantilever beam is a classical problem in solid mechanics
\cite{timoshenko-young-1968} that has been applied in studies of a wide
range of applications. These studies include the response of a
cantilever to a uniformly distributed or varying load
\cite{timoshenko-young-1968} or to a surrounding fluid flow
\cite{kamrin-rycroft-nave-2012, pozrikidis-2010, pozrikidis-2011,
  farjoun-schaeffer-2011, li-schulgasser-cederbaum-1995}.  For example,
Pozrikidis~\cite{pozrikidis-2010} studied shear flow over a periodic
array of cylindrical rods attached to a substrate in order to determine
the macroscopic slip velocity. He also computed the hydrodynamics load
exerted along the rod as well as estimating the flow-induced
deflection. Shortly afterwards, Pozrikidis~\cite{pozrikidis-2011}
studied shear flow past an elastic rod attached to a plane surface (as
well as a doubly-periodic array of such rods), motivated by the study of
biological flows involving ciliated surfaces. His mathematical framework
combined slender-body theory for computing the hydrodynamic load with
classical beam theory for computing the rod dynamics.  Small vibrations
of a thin flexible beam immersed in a laminar flow was studied by
Farjoun and Schaeffer~\cite{farjoun-schaeffer-2011}, where the authors
assumed that the dominant restoring force was due to tension caused by
shear stress.  Goza and Colonius~\cite{goza2017strongly} presented a
strongly coupled immersed boundary method for FSI problems involving
thin deforming bodies. The method was found to be stable for arbitrary
choices of solid to fluid mass ratios.

Because we are motivated by the study of problems involving deforming
biofilm layers, where the effects of porosity can be important, we will
also be investigating the behaviour of porous flexible cantilevers in
response to fluid flow. The closest work we have been able to identify
regarding porous beams is a mathematical model for an incompressible
poro-elastic beam~\cite{yang-wang-2007} in which the internal porous
fluid flow is confined to the axial direction of the deformed
beam. Other related studies of poroelastic beams can be found in
\cite{li-schulgasser-cederbaum-1995, Li-Schulgasser-Cederbaum-1998,yang-wang-2007,yang-wen-2010}, where again the beam was restricted to
be permeable in the axial direction only.

Our modelling approach is based on the immersed boundary or IB method,
which has been used in the study of problems in fluid-structure
interaction problems from biology in particular, but also in a variety
of engineering and other applications (see \cite{fauci-dillon-2006, peskin-2002} and references therein). The problem under consideration
here involves fluid interacting with rigid 1D structures (channel walls)
as well as a 2D deformable region (a beam, which can be either porous or
solid). Hence, our problem has a mixture of different solid geometries
for which the IB method is an ideal choice~\cite{peskin-1977, peskin-2002}.  Some researchers have applied the IB
method to simulate the dynamics of cilia which individually bear a
striking resemblance to a flexible cantilever beam (e.g.,
\cite{dillon-fauci-2000, dillon-etal-2007, schwartz-etal-1997}). The
issue of porous elastic boundaries has also been addressed in the IB
context, for example by Kim and Peskin \cite{kim-peskin-2006} who were
the first to incorporate porosity within the IB framework in a study of
parachute dynamics. Their approach to handling porous air vents at the
apex of the chute was to allow allow the normal velocity of the canopy
to differ from that of the fluid by an amount proportional to the normal
component of the boundary force.  Stockie \cite{stockie-2009} followed
Kim and Peskin's approach by incorporating porosity directly through
Darcy's law. We extended the ideas in \cite{kim-peskin-2006,
  stockie-2009} for a 1D porous membrane with pores directed normal to
the surface, to handle the case of solid porous 2D region.  We then
apply this approach to study deformation of a porous rectangular
cantilever beam with no restriction on the direction of the pores.  For
the general situation we consider here, we have not been able to find
any other numerical or experimental results with which we can draw
direct comparisons.

The organisation of this paper is as follows.  We start in
Section~\ref{sec:ib-method} by defining the problem geometry and list
the governing equations.  We also define the
IB force density that encompasses the influence of the
cantilever beam and bounding walls on the flow.  In Section
\ref{sec:solid_deform} we perform numerical simulations of an elastic,
cantilever beam and compare the results to experimental and analytical
results available in the literature.  Finally, in
Section~\ref{sec:porous}, we modify the governing equations to introduce
porosity into the beam and compare the results from numerical
simulations to the non-porous case.

\section{Immersed boundary method}
\label{sec:ib-method}

The term ``immersed boundary method'' is commonly used in the
literature to refer not only to a numerical method but also to the
underlying mathematical formulation~\cite{peskin-2002}.  In this
work, we only list the governing equations. The details of the governing equations as well as the numerical algorithm has been discussed  in \cite{ghosh2013immersed,ghosh2015numerical}.

\subsection{Governing equations}
\label{sec:model}

\begin{figure}[tbhp]
  \centering
\includegraphics[width=0.75\textwidth]{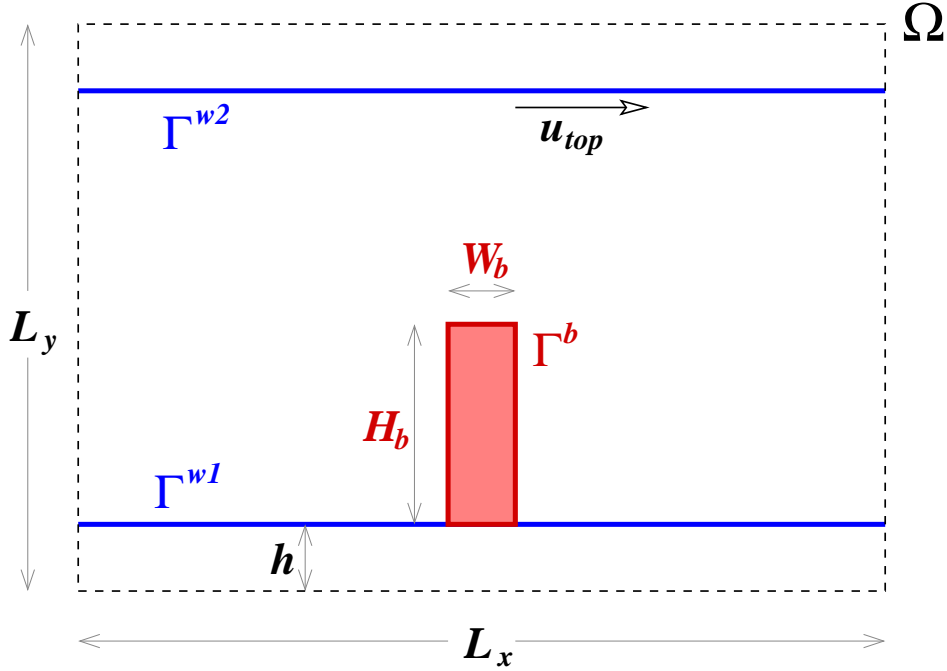}
\caption{A general elastic boundary $\Gamma = \mathop{\bigcup}_{i=1}^{3}
  \Gamma^i$ that consists of several possibly disconnected components is
  immersed within a doubly-periodic rectangular fluid domain $\Omega$.}
  \label{fig:domain}
\end{figure}

The IB formulation consists of a coupled system of
nonlinear integro-differential equations that describe both the force
generated by a solid, deformable, elastic body and the dynamics of a
surrounding incompressible, Newtonian fluid.  The method is capable of
handling immersed boundaries with a very general shape and
configuration, although  this study has been restricted  to the 2D geometry
pictured in Figure~\ref{fig:domain} wherein an initially rectangular
cantilever beam is placed inside a fluid-filled channel.  The channel
is embedded inside a larger, rectangular fluid domain $\Omega= [0,L_x]
\times [0,L_y]$ and is represented by two horizontal walls.  For
simplicity, we impose periodic boundary conditions in both the $x$- and
$y$-directions of the fluid domain, and the top and bottom channel walls are
horizontal lines separated a small distance $\dsep$ from
the top and bottom domain boundaries respectively.  The lower channel
wall $\Gamma^{w1}$ is fixed in space along $y=\dsep$, while the upper
wall $\Gamma^{w2}$ lies along $y=L_y-\dsep$ and moves to the right with
a given constant horizontal speed $\utop>0$.  In the absence of
obstructions in the channel, this setup would generate a
horizontal linear shear flow with shear rate $\gamma=\utop/(L_y-2\dsep)$.

The deformable cantilever beam $\Gamma^b$ is initially a rectangle with
thickness $\Wbeam$ and length $\Hbeam$, and has its lower end fixed to
the center of the bottom wall.  Note that the channel walls
$\Gamma^{w1}$ and $\Gamma^{w2}$ are taken to be idealized
one-dimensional structures of zero thickness, while the beam is treated
as a solid elastic structure, with a well-defined thickness.  The
details regarding the specification of the geometry and forces on the
walls and beam are delayed until Section~\ref{sec:force-density}.
The governing equations for this problem are:
\begin{itemize}
\item The motion of the fluid is governed by incompressible Navier-Stokes equations. The variables involved are: $\vu$ \bunits{cm/s}(fluid velocity), $p$ \bunits{g/cm\,s^2}(pressure), $\rho$ \bunits{g/cm^3} (density), $\mu$ \bunits{g/cm\,s}(dynamic viscosity), $\fsub{\vf}{IB}(\vq,t)$ \bunits{g/s^2} ( IB forcing term which captures the effect of the immersed elastic structure on the surrounding fluid)

\begin{gather}
  \rho \frac{\partial\vu}{\partial t} + \rho \vu\cdot \nabla \vu =
  \mu \nabla^2 \vu  - \nabla p + \fsub{\vf}{IB},
  \label{eq:ns-mom}
  \\
  \nabla \cdot \vu = 0. \label{eq:ns-inc}
\end{gather}
\item IB forcing term is calculated by the following equation:
\begin{gather}
  \fsub{\vf}{IB}(\vx,t) = \int_\Gamma
  \fsub{\vF}{IB}(\vq,t) \, \delta(\vx-\vX(\vq,t) )\,
  d\vq,
  \label{eq:ib-force}
\end{gather}
The variables involved are: $\fsub{\vF}{IB}(\vq,t)$ $\bunits{g/s^2}$(elastic force density) , $\Gamma$ (immersed structure),$\vX(\vq,t)$ ($\vX$ \bunits{cm})describes the location of the immersed structure and $\vq$ represents
dimensionless parameterization of points on $\Gamma$ ), $\vx=(x,y)$ \bunits{cm}( Fluid domain $\Omega$'s   Eulerian coordinates), $\delta(\vx)=\delta(x)\delta(y)$ ( 2D dirac delta function represented by cartesian product of two 1D dirac delta function).

\item The motion of the immersed structure is given by the following fibre evolution equation:
\begin{gather}
  \frac{\partial \vX}{\partial t} = \vu(\vX(\vq,t), t) =
  \int_{\Omega} \vu(\vx,t) \, \delta(\vx-\vX(\vq,t))\, d\vx,
  \label{eq:ib-velocity}
\end{gather}
The discrete version of the above equations are discussed in \cite{ghosh2013immersed,ghosh2015numerical}. In the next section, we discuss the discrete version of $\fsub{\vF}{IB}$.
\end{itemize}
\subsection{Discrete specification of the IB force density}
\label{sec:force-density}

We next specify the elastic IB forces that are generated by the channel
walls and cantilever beam pictured in Figure~\ref{fig:domain}.  Keep in
mind that because of the two-dimensional geometry, the equivalent
three-dimensional flow can be envisioned as extending to infinity in
both directions perpendicular to the $x,y$--plane, and hence the walls
are actually horizontal planar surfaces whereas the beam behaves as an
infinite cantilever plate.  We separate the IB force density
$\fsub{\vF}{IB}$ into the sum of three terms, $\fsub{\vF}{IB}=\Fwall +
\Fbeam + \Fattach$, where $\Fwall$ represents the force generated by the
channel walls, $\Fbeam$ is that generated by the beam, and $\Fattach$
derives from the attachment force between the wall and the bottom edge
of the beam.  Each of these forces is developed separately in the
following three sections.

\subsubsection{Force density for channel walls, $\Fwall$}
\label{sec:wall-force}

Each horizontal wall is discretized using a sequence of IB points that
are equally spaced in the fiber parameter $s$.  For example, along the
stationary bottom wall we define the initial wall point positions by
$\Xwallb_\ell = \left( \ell \hwall, \dsep \right)$, where $\hwall=L_x/N_w$ and
$\ell=1, 2, \dots, N_w$.  In the IB framework, the wall is actually
permitted to deviate slightly from its target configuration by
connecting each wall point to a fixed ``tether point'' (initially at the
same location) using a very stiff spring that exerts a force of the
form
\begin{gather}
  \Fwallb_\ell = \sigma_w (\Xwallb_\ell - \vX_\ell),
  \label{eq:fwall2}
\end{gather}
where $\sigma_w$ \bunits{g/cm\,s^2} is the wall spring stiffness and
$\vX_\ell$ is the location of the moving IB point.  Any deviation of the wall
point from corresponding the tether point location generates a force
that brings the IB point back towards the tether point, so if the value of
$\sigma_w$ is large then  the wall points can be treated as
rigid structure.  The tether points do not generate
any force  but only serve to determine target locations for
the moving IB points.

The given horizontal velocity of the top wall points is easily
incorporated in the above framework by simply specifying a given motion
for the tether points, so that the IB force density becomes
\begin{gather}
  \Fwallt_\ell = \sigma_w (\Xwallt_\ell(t) - \vX_\ell),
  \label{eq:fwall1}
\end{gather}
where $\Xwallt_\ell(t) = \big(\operatorname{mod} \left(\ell \hwall + \utop
  t, L_x\big), L_y-h \right)$ and the ``modulo $L_x$'' operation
enforces the periodic boundary conditions in $x$.  The total force
density generated by the two channel walls may then be written as
\begin{gather}
  \Fwall = \sum_{\ell=1}^{N_w} (\Fwallt_\ell + \Fwallb_\ell).
\end{gather}

\subsubsection{Force density for cantilever beam, $\Fbeam$}
\label{sec:beam-force}

\begin{figure}[htbp]
  \centering
  \begin{tabular}{ccc}
    (a) Rectangular beam & & (b) Smoothed beam\\
    (731 points, 1304 triangles) & & (743 points, 1320 triangles)\\
    \includegraphics[height=0.36\textheight]{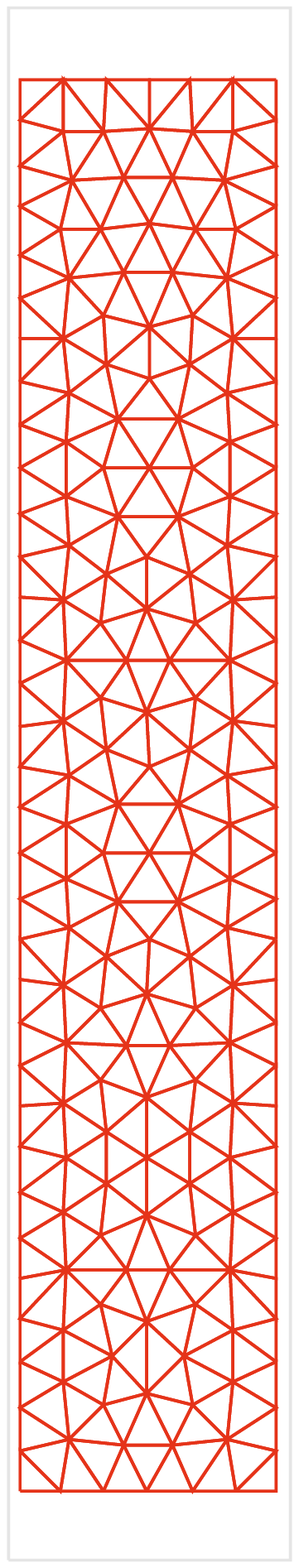}
    & \qquad\qquad\qquad &
    \includegraphics[height=0.36\textheight]{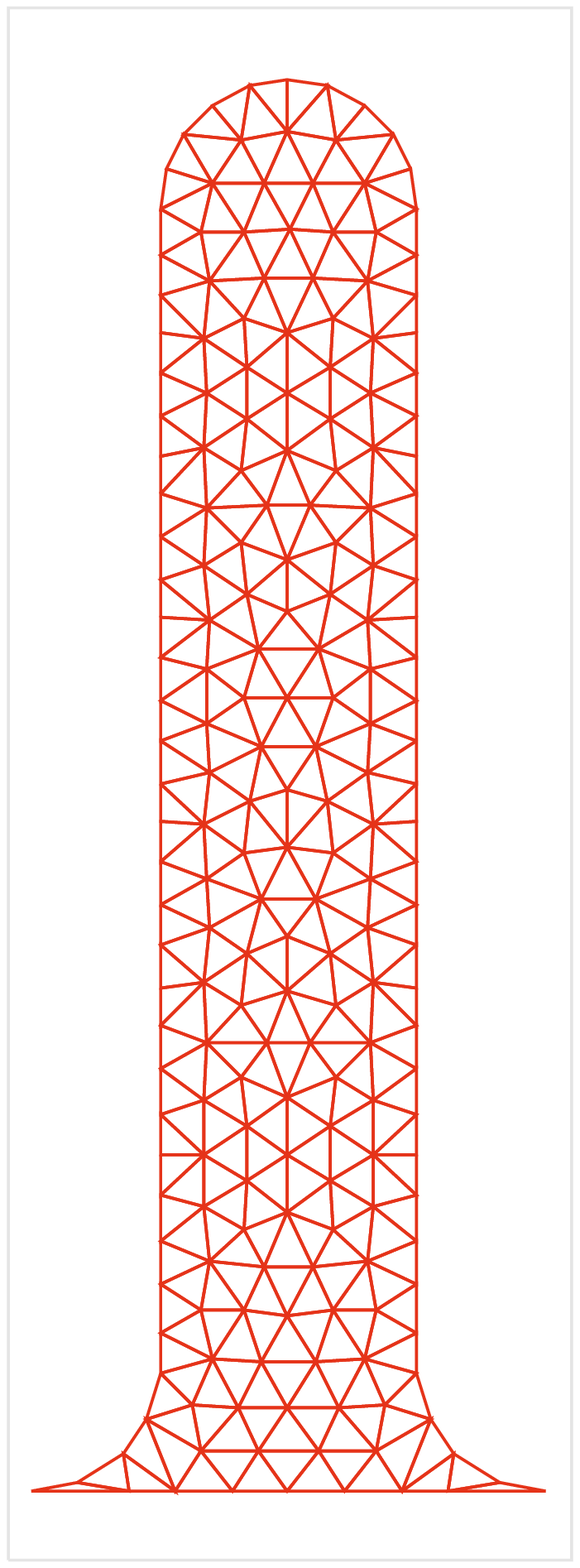}
  \end{tabular}
  \caption{Unstructured triangular mesh for the two beam shapes,
    generated by {\tt distmesh2d}.  Both have the same height and width,
    and the fillets for the smoothed beam are constructed in such a way
    that the areas of the two shapes are equal.}
  \label{fig:dist_mesh}
\end{figure}

The cantilever beam is discretized using a collection of $N_b$
Lagrangian points that lie on its circumference as well as distributed
throughout the interior of the beam.  We employ the unstructured triangular mesh
generator {\tt DistMesh}~\cite{persson-strang-2004} (implemented in
Matlab) to generate an approximately uniform triangulation for the beam
such as that depicted in Figure~\ref{fig:dist_mesh}a.  The nodes of the
triangulation are the IB points $\vX_\ell$, for $\ell=1,2,\dots, N_b$ and
a network of springs is defined by the edges of the triangles that acts
to maintain the shape of the beam.
We generate the triangulation using the function {\tt distmesh2d} with
the ``scaled edge length function'' {\tt huniform}, which is a built-in
function that aims to find a mesh that is as uniform as possible.  The
  ``initial edge length'' parameter has been assigned equal to
$\min(\dx,\dy)/3$, which ensures that in practice the mesh satisfies the
following constraint
\begin{gather}
  \max_{k,\ell} |\vX_k-\vX_\ell| < \frac{1}{2}\,\min(\dx,\dy),
  \label{eq:leak-constraint}
\end{gather}
and hence prevents leakage of fluid through the IB that
can occur if IB points become too widely separated~\cite{peskin-2002}.
We will also perform numerical simulations on the ``smoothed beam''
shape shown in Figure~\ref{fig:dist_mesh}b, where the rectangular
corners are replaced by fillets or chamfers -- the reasons for the
choice of this smoothed shape will be explained later in
Section~\ref{sec:vary-utop}.

We have defined  the spring forces acting on the network,  based on
the model of Alpkvist and Klapper for viscoelastic biofilm
structures~\cite{alpkvist-klapper-2007}.  The vector joining two IB points labeled
$\ell$ and $m$ is denoted by  $\vd_{\ell,m}(t) =
\vX_\ell(t) - \vX_m(t)$ , and the corresponding distance by $d_{\ell,m}(t) = |\vd_{\ell,m}(t)|$. The network of springs is initially
assumed to be in equilibrium (i.e., all spring forces are in balance) so that resting length  of each spring is equal to its initial length $d_{\ell,m}(0)$. An incidence matrix whose entries
$\mathbb{I}_{\ell,m}$ are either 1 or 0 depending on whether or not the IB
points labelled $\ell$ and $m$ are connected to each other is represented by
 $\mathbb{I}$.  Then the net force
density acting on the $\ell^{th}$ IB point in the network is given by
\begin{gather}
  \Fbeam_\ell =
  \sum_{m=1}^{N_b} \sigma_b
  \mathbb{I}_{\ell,m} \frac{\vd_{\ell,m}}{d_{\ell,m}}
  (d_{\ell,m}(0) - d_{\ell,m}),
  \label{eq:fbeam1}
\end{gather}
where the summation is done over those nodes m of the network for which $\vX_m$  is connected to $\vX_\ell$.
The spring stiffness $\sigma_b$ \bunits{g/cm\,s^2} is assumed to be  constant for all
network connections.  The total elastic force density will be generated by all
IB points of  the beam and will be represented  by the expression:
\begin{gather}
  \Fbeam = \sum_{\ell=1}^{N_b}\Fbeam_\ell.
  \label{eq:fbeam}
\end{gather}
%

\subsubsection{Force density for attachment between beam and bottom wall, $\Fattach$}

In order to hold the lower end of the cantilever beam stationary and
coincident with the channel wall, we impose an additional attachment
force that connects the points at the base of the beam to the fixed
tether points on the bottom wall.  The wall and beam points are
initially collocated and each pair of points is connected by a very
stiff spring.  Denote by ${\cal A}$ the attachment point index set which
consists of all pairs of IB point indices $(\ell,m)$ for which $\ell$
corresponds to the bottom wall tether point that is linked to the beam
IB point $m$.  Then the force density arising from the spring joining
these two points is
\begin{gather}
  \Fattach_{\ell,m} = - \sigma_a \, \vd_{\ell,m},
  \label{eq:beq4}
\end{gather}
where $\vd_{\ell,m} = \Xwallb_\ell - \vX_m(t)$ is the vector connecting
the wall tether point with fixed location $\Xwallb_\ell$ and the beam
point at $\vX_m(t)$.  Note that the springs in this case have a zero
resting length, so that the IB forces act to keep the point pairs in the
same location.  The spring constant $\sigma_a$ is chosen equal to that
of the wall--tether connections so that $\sigma_a = \sigma_w \gg \sigma
_b$.  The total wall--beam attachment force can then be written as
\begin{gather}
  \Fattach = \sum_{(\ell,m)\in {\cal A}}  \Fattach_{\ell,m}.
\end{gather}

\section{Numerical simulations of a solid beam}
\label{sec:solid_deform}

Our numerical simulations presented in this section consist of a
parametric study in which we vary values of the beam spring stiffness
$\sigma_b$, beam length $\Hbeam$, and top wall velocity $\utop$, the
latter of which is related to shear rate via $\gamma =
\utop/(L_y-2\dsep)$.  We start with a ``base case'' corresponding to a
beam having dimensions $\Hbeam=0.0077$ and $\Wbeam=0.0014$.  The fluid
domain is a square of size $L_x=L_y=0.03$, and the channel walls are
located a distance $\dsep=0.00328$ from the boundaries, which means that
the channel width is $L_y-2\dsep=0.0234$.  This choice of geometry is
motivated by one of the test cases considered by Alpkvist and Klapper in
\cite{alpkvist-klapper-2007}, and yields a beam aspect ratio
$\Hbeam/\Wbeam\approx 5$ that is large enough to satisfy the thin-beam
assumption in the linear theory.  This beam length is also small enough
(roughly one-third of the channel width) that it doesn't significantly
hinder the bulk flow through the channel.  The base value for the top
wall velocity is $\utop=0.02$, while the spring stiffness values are
$\sigma_b=560$ and $\sigma_w=\sigma_a=1000$.  For the numerical
discretization, we take a fluid grid with $N_x=N_y=64$ grid points in
each direction, and choose $N_b=731$ and $N_w=210$ in order to satisfy
the constraint \en{leak-constraint} on the spacing between IB points for
the beam and walls.  The base case values of all parameters are
summarized in Table~\ref{tab:parameter}.

\begin{figure}[tbhp]
  \centering
  \begin{tabular}{ccc}
    (a) & & (b) \\
    \includegraphics[trim=70 0 70 0,clip,width=0.47\textwidth]{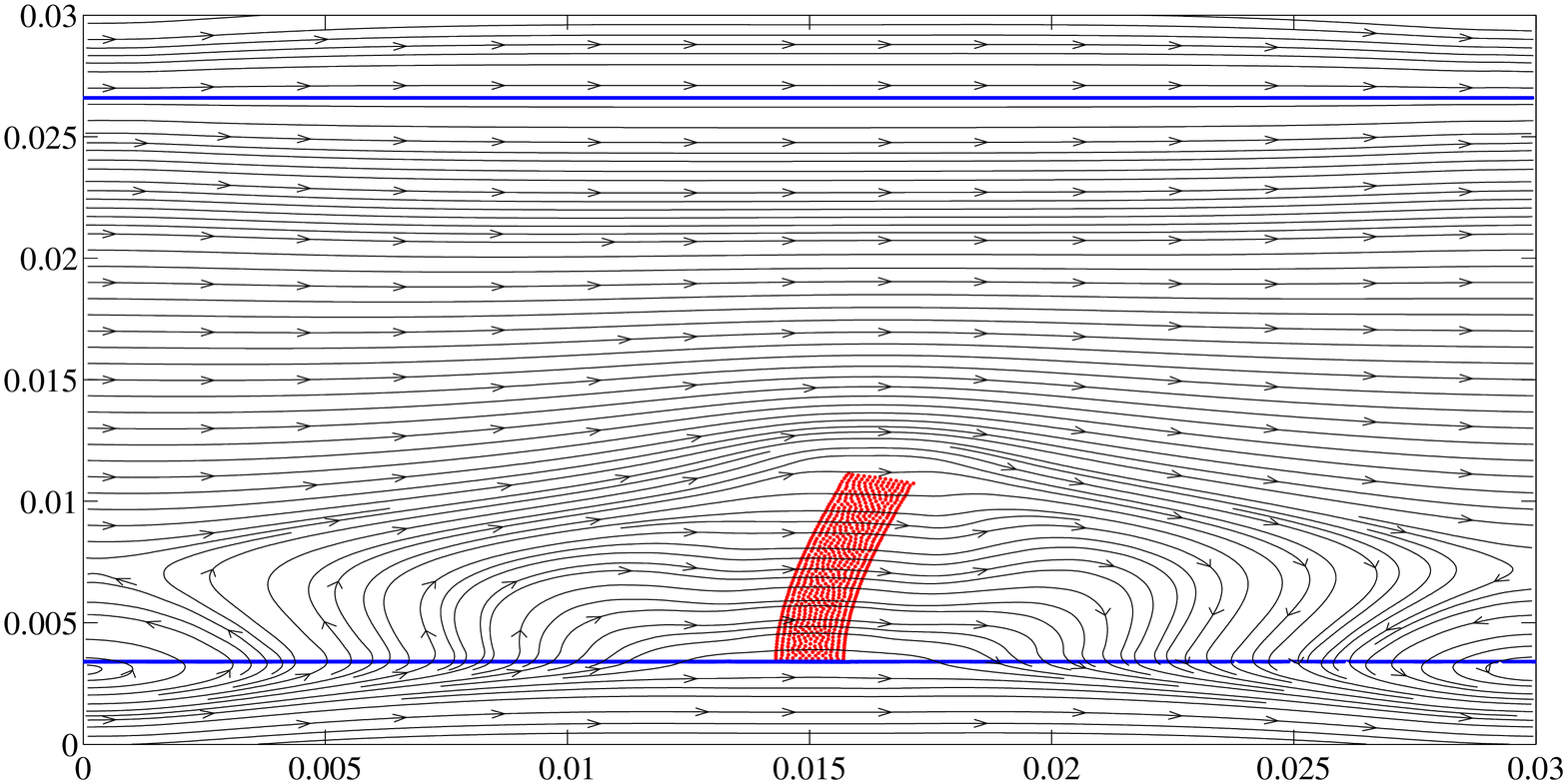}
    & &
    \includegraphics[trim=70 0 70 0,clip,width=0.47\textwidth]{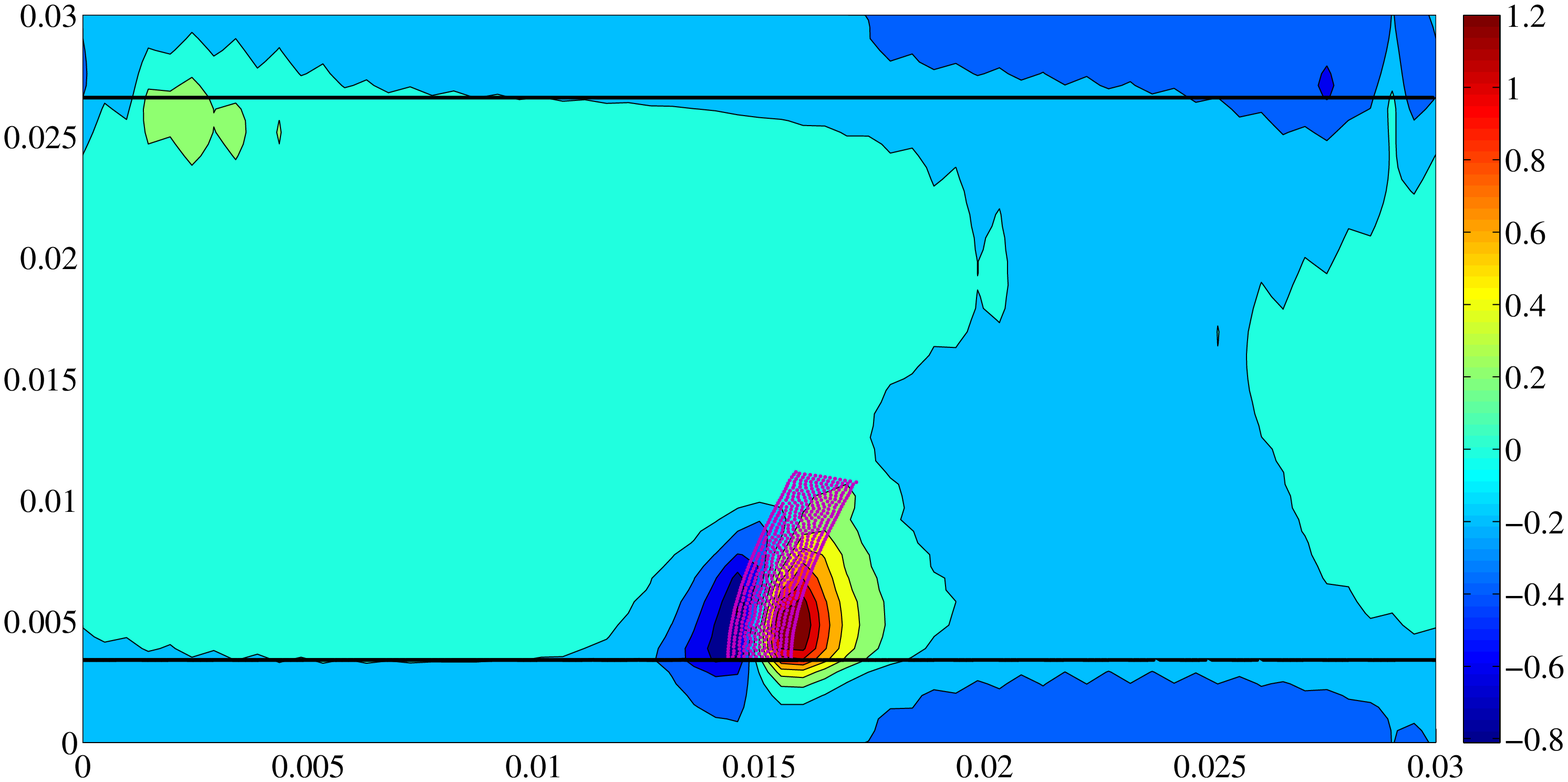}
  \end{tabular}
  \caption{Flow streamlines (left) and pressure contours (right) for the
    base case simulation at steady state.}
  \label{fig:base_case}
\end{figure}

As parameters are varied from the base case, we focus on the deflection
of the beam from its vertical stress-free state and also the changes in
the flow structure.  In all cases, the fluid velocity is initialized to
zero and the speed of the top wall is ramped up linearly from 0 to
$\utop$ over the time interval $[0,0.5]$.  As the fluid within the
channel accelerates and begins to move toward the right, the beam
responds by bending downward and to the right as pictured in
Figure~\ref{fig:base_case}.  Because the beam is solid, the flow
deflects around the beam, generating a fairly complex flow structure
that features recirculation zones in the area near the bottom wall
(refer to the streamline patterns in Figure~\ref{fig:base_case}a).  The
beam continues to deform in response to the flow until an equilibrium is
attained in which the fluid force acting on the beam and the elastic
bending force from the deformed beam are equal.  A typical computation
requires on the order of 5 to 10~\units{s} for the beam to reach its
steady state.

%

\leavethisout{
  \begin{note}[Geometrical parameters]
    We should still draw some comparison of $L_x, L_y, h$ to
    Alpkvist-Klapper, and also add a justification for the numerical
    parameters $N_x, N_y, N_b, N_w$.
  \end{note}
}

We remark that during the course of these rectangular beam simulations,
especially for cases with large deflection, we sometimes observe large
non-physical irregularities in the shape of the beam in the vicinity of
the free end and the attachment at the lower wall especially at high
shear rates.  For this reason, we also perform a series of simulations
with a smoothed beam shape for which the corners at the free end and the
beam--wall anchor points are rounded using a fillet or chamfer (refer to
Fig.~\ref{fig:dist_mesh}b).  These results are presented in
Section~\ref{sec:smoothed-beam}, and we will see that they exhibit more
reasonable deformations near the corners.

\leavethisout{
  A typical plot of the solution showing the flow
  streamlines and the deformed beam shape is given in
  Figure~\ref{fig:flow_profile}.
  \begin{figure}[tbhp]
    \centering
    \includegraphics[width=0.75\textwidth]{streamline_beam_modified} \\
    \caption{A typical flow pattern for the rectangular cantilever
      beam.}
    \label{fig:flow_profile}
  \end{figure}
}

\begin{table}
  \begin{center}
    \begin{tabular}{|c|c|c|c|}
      \hline
      Parameter & Description & Value & Units\\
      \hline\hline
      $L_x$      & Domain width               & 0.03   & $\units{cm}$\\
      $L_y$      & Domain height              & 0.03   & $\units{cm}$\\
      $\dsep$    & Wall separation distance   & 0.0033 & $\units{cm}$\\
      $\Hbeam$   & Beam length                & 0.0077 & $\units{cm}$\\
      $\Wbeam$   & Beam width                 & 0.0014 & $\units{cm}$\\
      $\sigma_b$ & Beam spring stiffness      & 560    & $\units{g/cm\,s^2}$\\
      $\sigma_w$ & Wall spring stiffness      & 1000   & $\units{g/cm\,s^2}$\\
      $\sigma_a$ & Attachment spring stiffness& 1000   & $\units{g/cm\,s^2}$\\
      $\utop$    & Top wall velocity          & 0.02   & $\units{cm/s}$\\
      $\rho$     & Water density              & 1.0    & $\units{g/cm^3}$\\
      $\mu$      & Water viscosity            & 0.01   & $\units{g/cm\,s}$\\
      $\Reynolds$& Reynolds number $=\rho\utop\Hbeam/\mu$ & 0.0154 &\\
      $N_x$, $N_y$ & Number of fluid grid points & 64  &\\
      $N_b$      & Number of beam IB points   & 731    &\\
      $N_w$      & Number of wall IB points   & 210    &\\
      $\dt$      & Time step                  & $10^{-5}$     & $\units{s}$\\
      \hline
    \end{tabular}
  \end{center}
  \caption{``Base case'' parameter values for the solid elastic beam
    problem.}
  \label{tab:parameter}
\end{table}

\subsection{Euler-Bernoulli beam theory}
\label{sec:linear-theory}

Before reporting the results of our numerical simulations, we briefly
summarize some key results from the Euler-Bernoulli beam theory that
pertain to small deflections of a thin cantilever beam.  In this case
the steady-state configuration of the beam can be desribed by a
fourth-order ordinary differential equation~\cite{timoshenko-young-1968}
\begin{gather}
  EI \frac{d^4 \xi}{dy^4} = q,
  \label{eq:beam4}
\end{gather}
where $\xi(y)$ is the horizontal deflection from the vertical equilibrium
state at location $y$ along the beam, and $q(y)$ is a given load
distribution (in units of force per unit length, \units{dyne/cm} or
\units{g/s^2}).  The parameter $EI$ is the flexural rigidity of the
beam, which is the product of the Young's modulus $E$ \bunits{g/cm\,s^2}
and the second moment of area $I$ \bunits{cm^4}.  The linear theory is
derived based on the assumptions that the beam is very thin ($\Wbeam \ll
\Hbeam$) and that the deflection $\deflect$ is small relative to the
length of the beam ($\deflect / \Hbeam \ll 1$).

In particular, we list well-known analytical solutions for the maximum (tip)
deflection of the beam in response to two types of applied load:  (a) a constant load $q_0$
\bunits{dyne/cm} distributed across the entire length of the beam : $\displaystyle \deflect = \frac{q_0\Hbeam^4}{8EI}$ and
(b) a linearly-varying load that decreases from a maximum $q_0$ at the
wall to zero at the tip :$\displaystyle \deflect = \frac{q_0\Hbeam^4}{30EI}$ .  The tip deflection in all two cases varies according
to $\deflect\propto \Hbeam^m/EI$, with exponent $m=4$.  These
expressions will be compared qualitatively to the simulated deflections in
Section~\ref{sec:vary-sigmab}.

\subsection{Dependence of deflection on IB spring stiffness, $\sigma_b$}
\label{sec:vary-sigmab}

We begin by investigating the effect of changing the spring stiffness
$\sigma_b$ for links in the beam triangulation, which in turn determines
the effective flexural rigidity of the overall beam structure.  We note
that the beam dimensions $\Hbeam=0.0077$ and $\Wbeam=0.0014$
corresponding to an aspect ratio of $\Hbeam/\Wbeam \approx 5$, which is
consistent with the thin beam assumption in the linear theory.

\begin{figure}[tbhp]
  \centering
  (a) Deflection versus time. \\
  \includegraphics[width=0.75\textwidth]{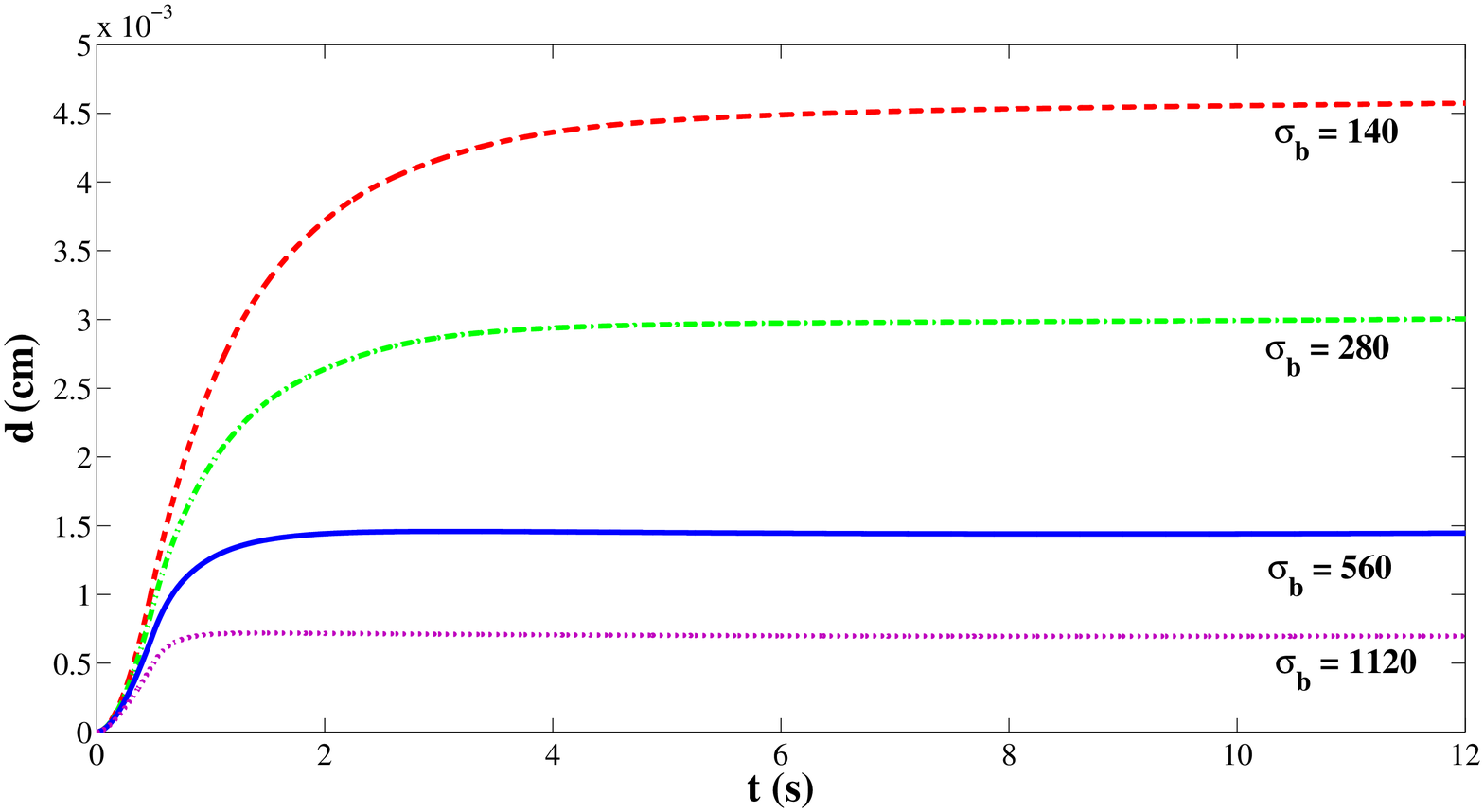}\\
  (b) Final deflection versus $\sigma_b$.\\
  \includegraphics[width=0.75\textwidth]{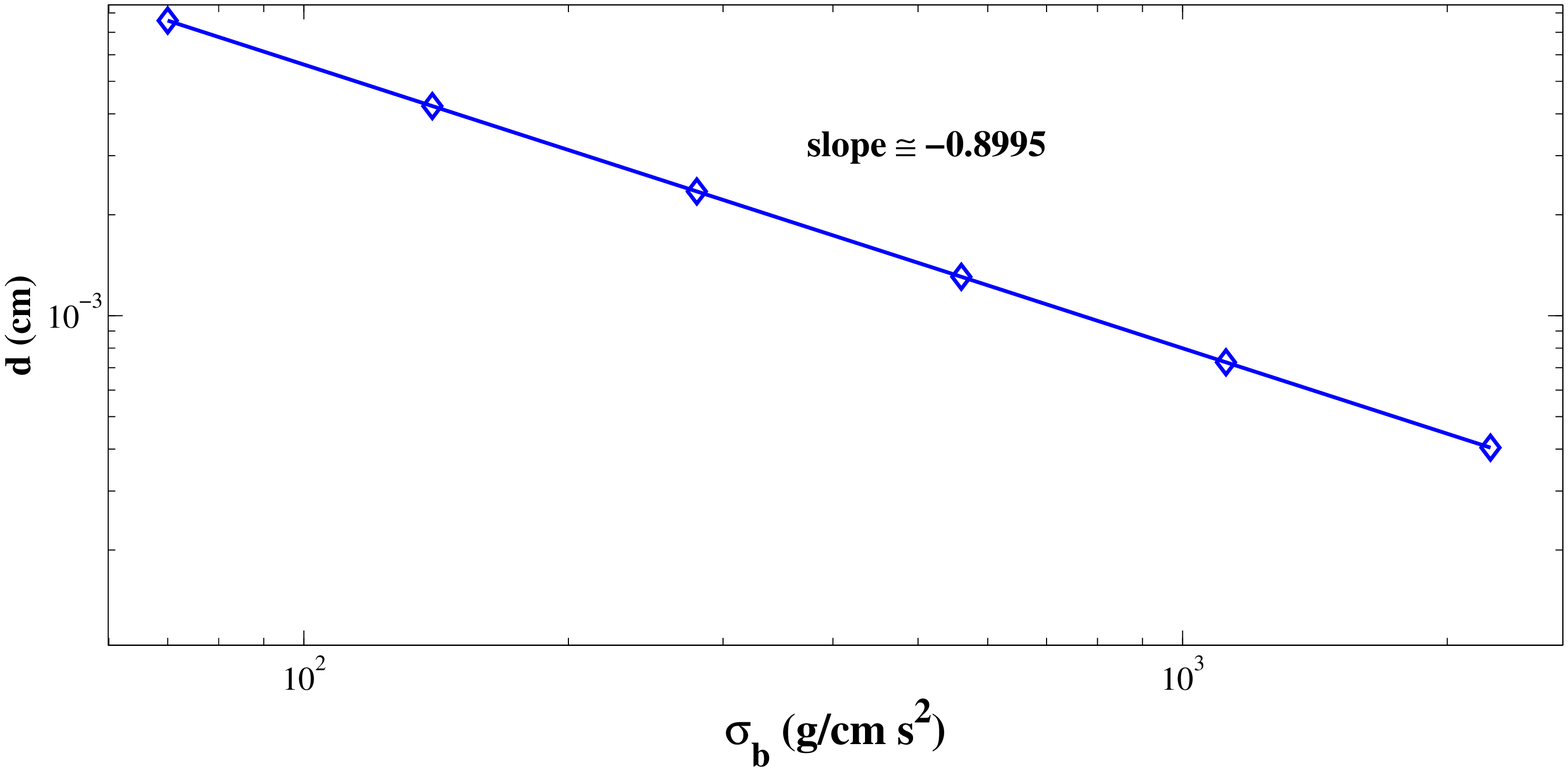}
  \caption{Deflection of the solid cantilever beam as $\sigma_b$ is
    varied, with $\Hbeam = 0.0077$ and $\utop=0.02$.  (a) Tip deflection
    versus time corresponding to $\sigma_b\in[70,2240]$.  (b) The final
    (steady-state) tip deflection is plotted against $\sigma_b$ with
    diamond points (\textcolor[rgb]{0.00,0.00,1.00}{$\diamond$}) on top
    of which is superimposed a solid line corresponding to the linear
    least squares fit (normalized RMS error $=$ 0.055).}
  \label{fig:deflect_cant}
\end{figure}

Figure~\ref{fig:deflect_cant}a depicts the time evolution of the beam
tip deflection for values of $\sigma_b= 140, 280,  560, 1120$, where
deflection $\deflect$ is measured in the $x$-direction from the vertical
equilibrium configuration.  The steady-state deflection values vary from
roughly 10\%\ of the beam length for the stiffest beam up to 65\%\ for
the smallest value of $\sigma_b=70$.  It seems reasonable therefore to
expect that the larger $\sigma_b$ simulations will fall within the small
deflection regime, while the smaller $\sigma_b$ results might not.
Nonetheless, all curves exhibit a similar qualitative behaviour in that
the shear flow bends the beam away toward the right and then gradually
equilibrates at some maximum deflection.  As expected, the stiffer beams
have a smaller deflection and also reach their equilibrium state over a
shorter time period.

\leavethisout{
  \begin{note}[Extra streamline plot]
    It might be nice to have a streamline plot in this section
    corresponding to the linear deflection regime.  I suspect that the
    deflection is not even visible in this case, but this would be a nice
    complement to the other figures where the deflection is much larger.
  \end{note}
  \begin{note}[Addition to Figure~\ref{fig:deflect_cant}a]
    It looks like there is a local maximum in most (all?) of these curves.
    So it would be helpful to indicate with a point the location/time at
    which the maximum deflection occurs in each case.  If there is no
    local maximum for (iv), then instead we could plot the point where the
    deflection comes within some small epsilon distance from the
    equilibrium state.
  \end{note}
}

Figure~\ref{fig:deflect_cant}b plots the steady-state deflection
$\deflect$ for all simulations versus $\sigma_b$, along with a number of
additional results for $\sigma_b$ in the interval $[70,2240]$ that fill
out the range.  The data points are approximated very well by a straight
line on a log--log scale; indeed, a least squares fit yields a line with
with slope $=-0.8995$ which suggests that the deflection is roughly
inversely proportional to $\sigma_b$.  We can compare these results to
the linear beam theory in Section~\ref{sec:linear-theory} which predicts
that the deflection is inversely proportional to the flexural rigidity,
$EI$, where $E$ is Young's modulus, $I$ is the area moment of inertia.
Assuming that the linear theory holds here (thin beam, small deflection)
our computational results suggest that the effective flexural rigidity
of our triangulated spring link structure is directly proportional to
the spring stiffness, or in other words that $EI\propto\sigma_b$.

\leavethisout{
  \begin{note}[Choice of $\utop$]
    I understand that you've chosen $\utop=0.00075$ so that the beam is
    within the linear deflection regime.  However, it is puzzling that
    this is chosen so far outside the range of $\utop$ values that you
    list for the parameter study in Table~\ref{tab:parameter}.  It would
    be nice if you could either add $\utop=0.00075$ to your parameter
    study, or else redo all of the simulations in this section at
    $\utop=0.003$.  The latter is obviously a lot more work.
    Regardless, I think it would be nice to extend this plot a bit
    further to larger deflections which would mean $\sigma_b<100$.  Why?
    Well, a deflection that is at most 1.4\%\ of the beam length is
    obviously in the linear regime; however, it's a bit too small and so
    pushing $d/H_b$ up to 5 or even 10\%\ would be a bit more realistic,
    and might even allow us to make a comment about when the linearity
    assumption breaks down.  But it could be that reducing
    $\sigma_b<100$ is not possible and that instead we'll have to take a
    larger $\utop$.  Before you start any new simulations, please let me
    know what you think about this, and whether or not you already have
    some other simulations that could be added here.
  \end{note}
}

\subsection{Dependence of deflection on beam length, $\Hbeam$}
\label{sec:vary-hbeam}

\begin{figure}[tbhp]
  \centering
  \includegraphics[width=0.75\columnwidth]{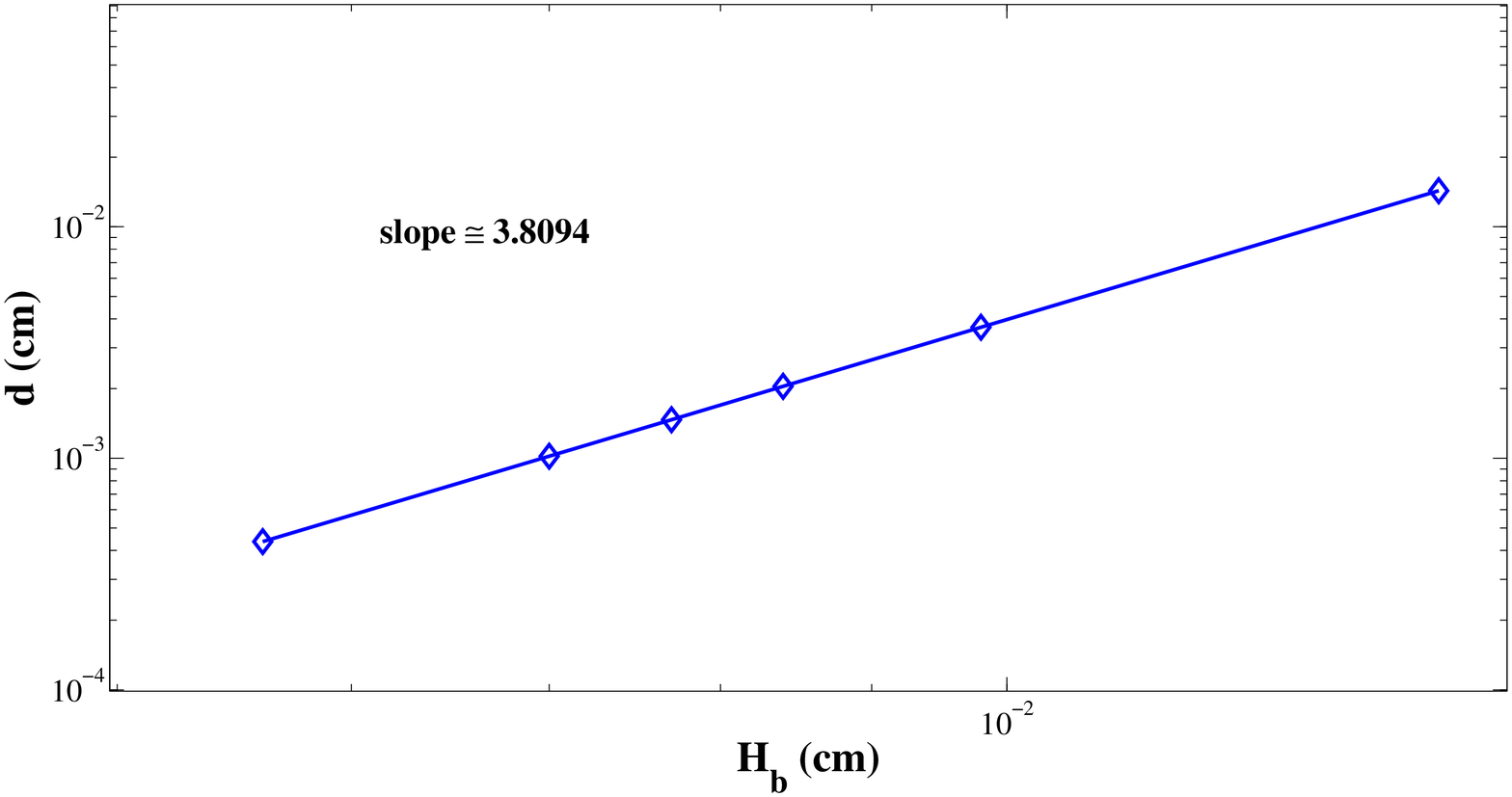}
  \caption{Plot of $\Hbeam$ and final deflection $\deflect$ for the
    non-porous cantilever. Parameters: $\sigma_b = 560 \;
    \units{g/cm\,s^2}$, $\utop= 0.02\; \units{cm/s}$. Computed results
    are shown using diamond points (\textcolor[rgb]{0.00,0.00,1.00}{$\diamond$})
    on top of which is superimposed a solid line corresponding to the linear
    least squares fit (normalized RMS error $=$ 0.0432).}
  \label{fig:def_vs_length_canti}
\end{figure}

Next we perform simulations with different values of the beam length
$\Hbeam = 0.0056$, 0.007, 0.0077, 0.0084, 0.0098, 0.014 (corresponding
to aspect ratios of $\Hbeam/\Wbeam = 4$, 5, 5.5, 6, 7, 10 respectively)
while holding $\sigma_b=560$ and $\utop=0.02$ constant.  The computed
results in Figure~\ref{fig:def_vs_length_canti} show that the tip
deflection $d$ increases with $\Hbeam$, which is to be expected since a
longer beam is clearly more flexible.  Furthermore, the dependence of
$d$ on $\Hbeam$ is again roughly linear on a log-log scale which denotes
a power law relationship, and a least squares fit suggests that
$\deflect \propto \Hbeam^{3.8094}$.  This power law exponent is very
close to the value 4 predicted by the linear beam theory for the cases
when the load force is constant or linearly-varying along the length.

\leavethisout{
  \begin{note}[More confusion over parameters]
    Please let me know what the beam length values should be here.  Why
    are the parameter values for this case different from the previous
    one?  Why would we use 280 instead of 260 for $\sigma_b$ (or do I
    have the value 260 wrong in the previous section)?  My confusion
    grows \dots\ And $\utop=0.000428$ doesn't match any of the values we
    used earlier.  Why?  We should really be picking a ``base'' set of
    parameters and then varying one parameter in each section from this
    base case.
  \end{note}
}

\subsection{Dependence of deflection on shear velocity, $\utop$}
\label{sec:vary-utop}

\begin{figure}[tbhp]
  \centering
  \begin{tabular}{m{0.47\textwidth}@{}m{0.47\textwidth}}
    (a) $\utop=0.003$ & \\
    \includegraphics[width=0.5\textwidth,clip]{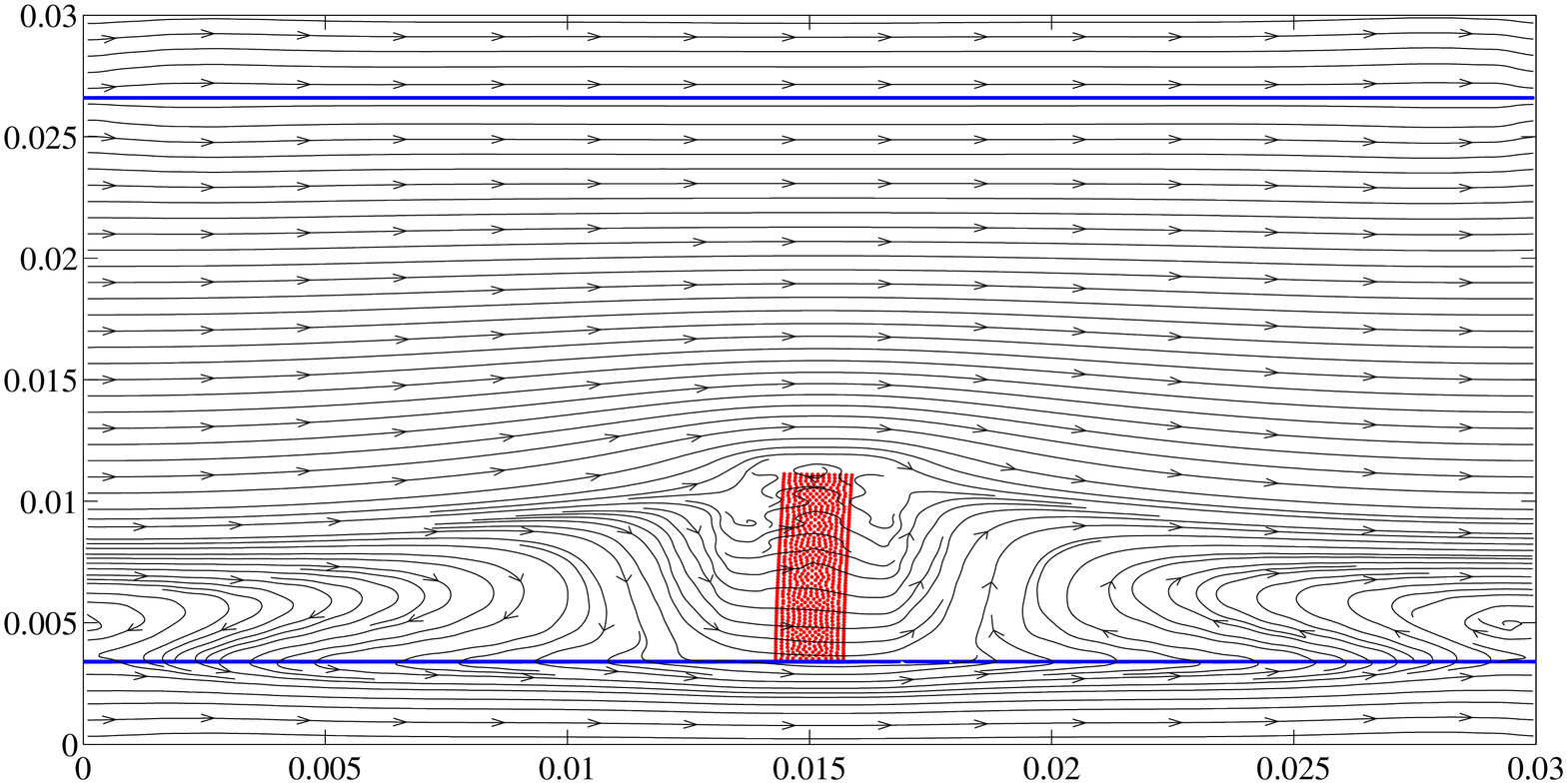}
    &
    \includegraphics[width=0.5\textwidth,clip]{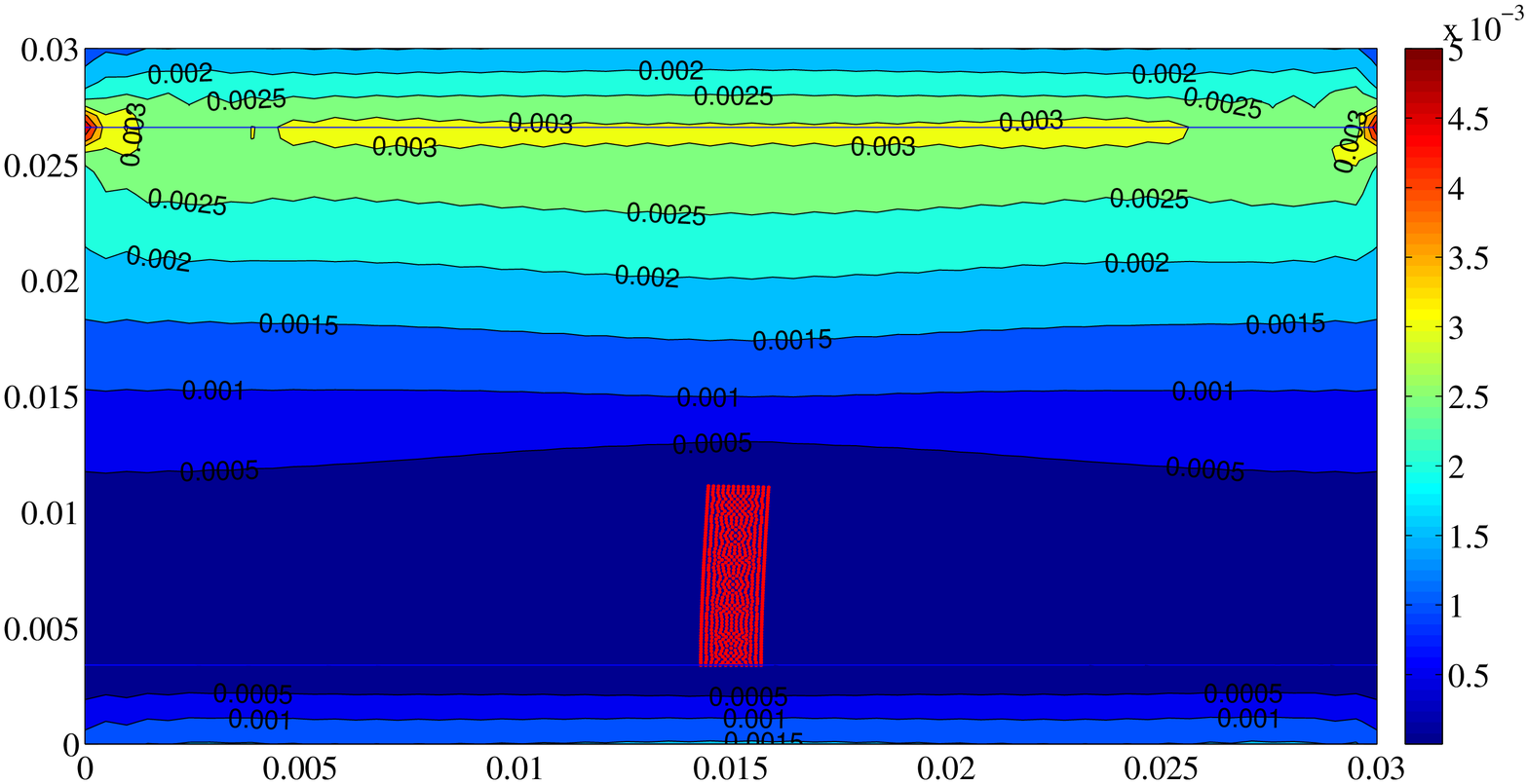}
    \\
    (b) $\utop=0.01$ & \\
    \includegraphics[width=0.5\textwidth,clip]{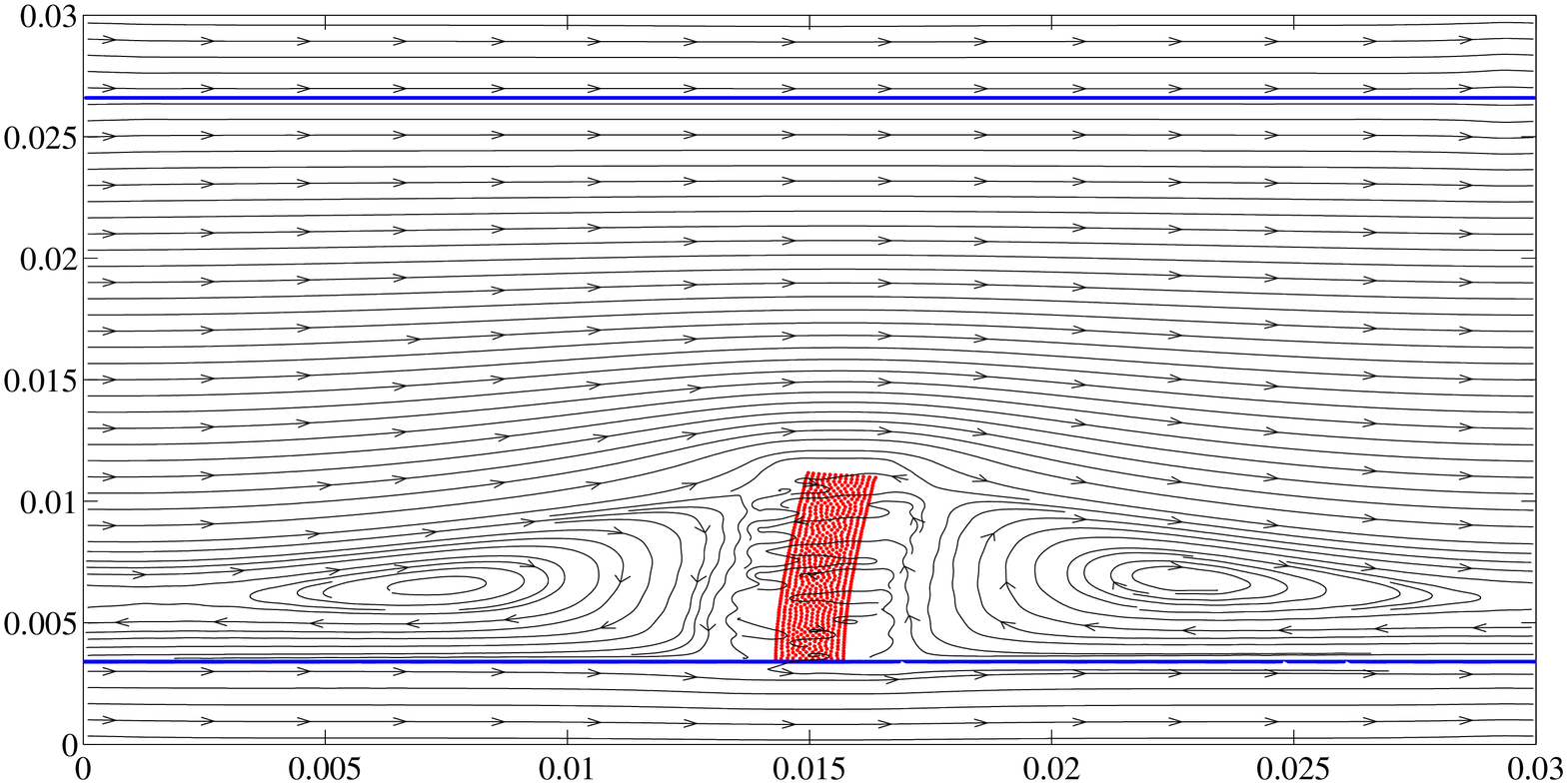}
    &
    \includegraphics[width=0.5\textwidth,clip]{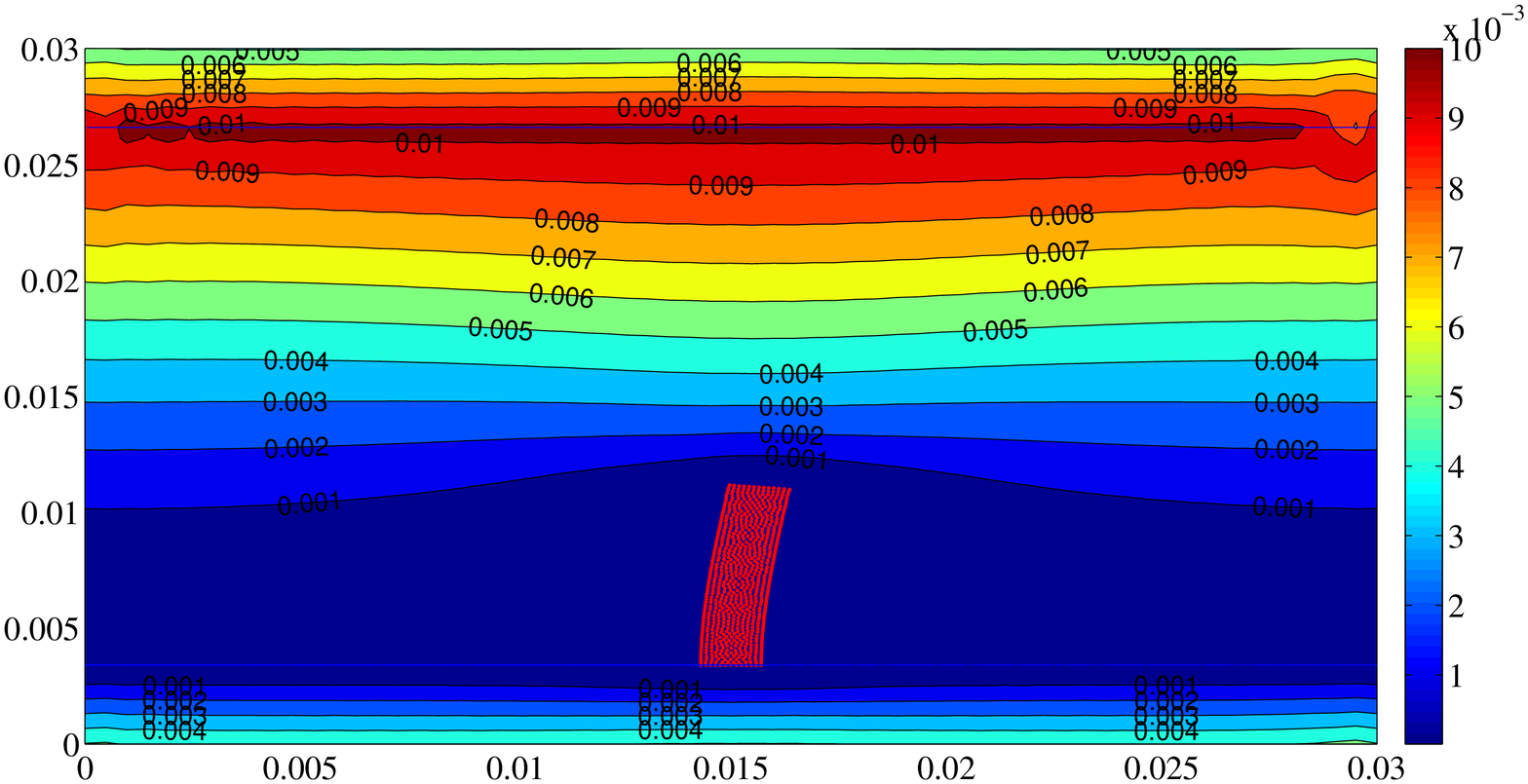}
    \\
    (c) $\utop=0.02$ (base case) & \\
    \includegraphics[width=0.5\textwidth,clip]{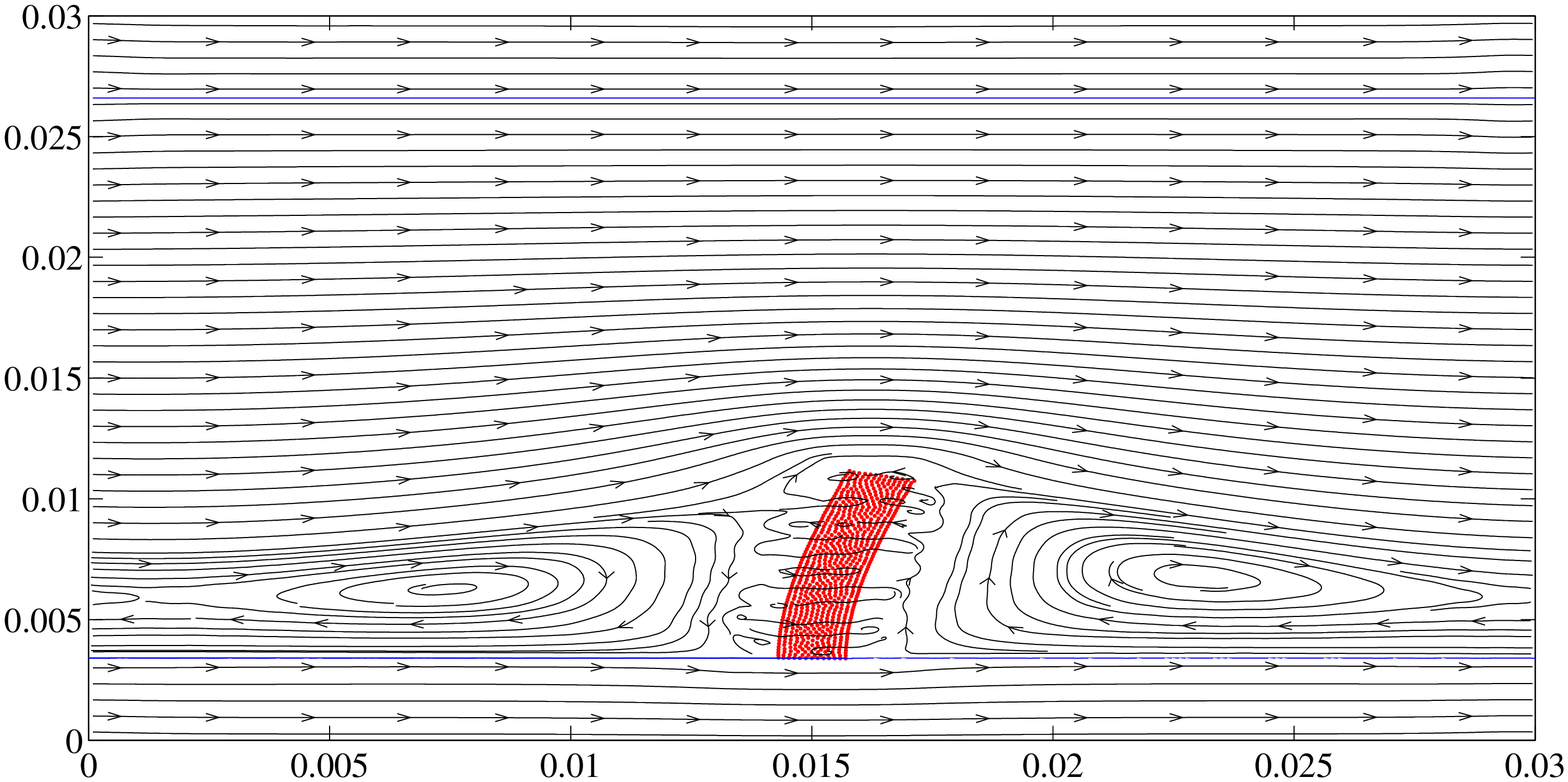}
    &
    \includegraphics[width=0.5\textwidth,clip]{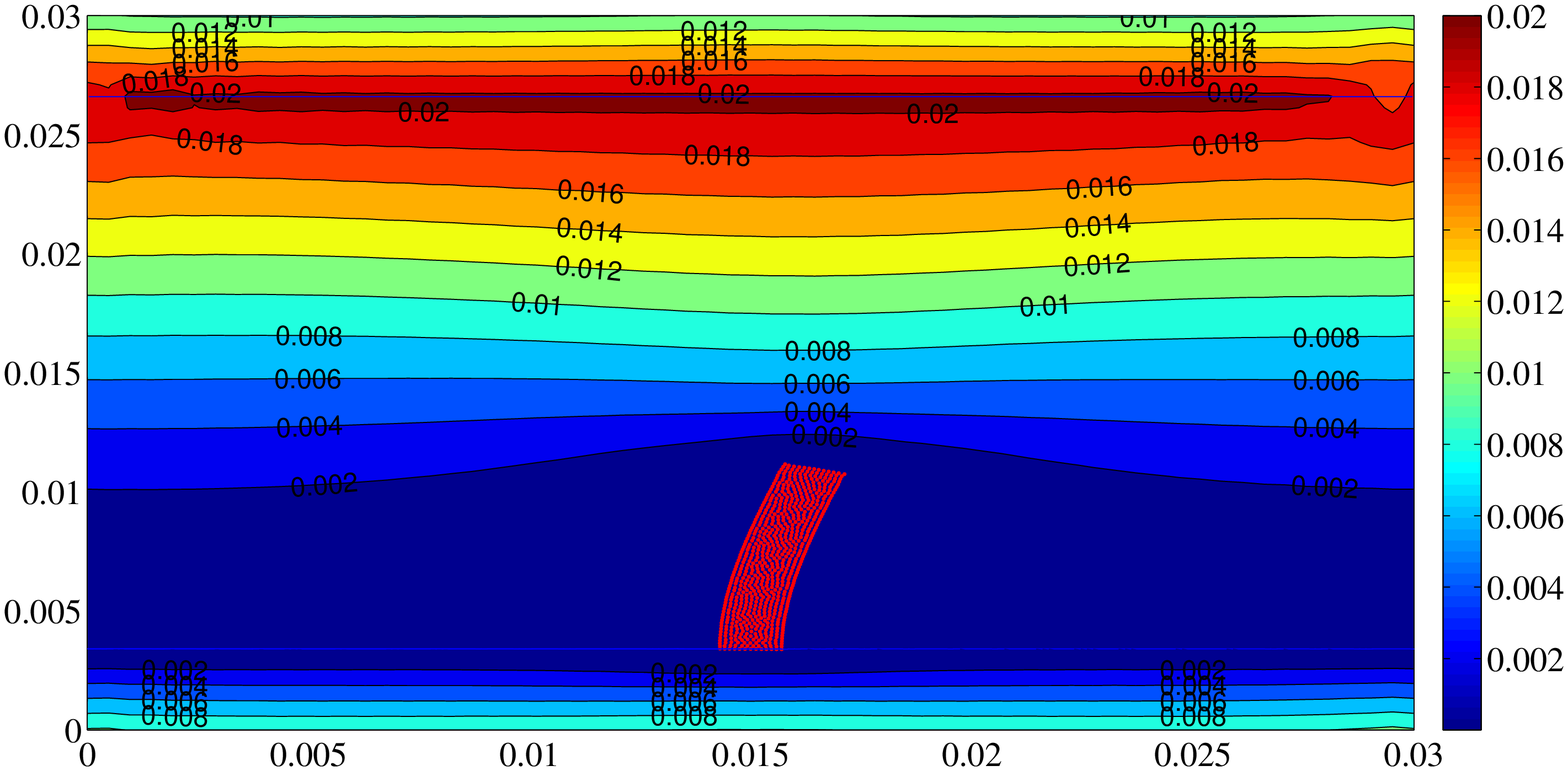}
    \\
    (d) $\utop=0.04$ & \\
    \includegraphics[width=0.5\textwidth]{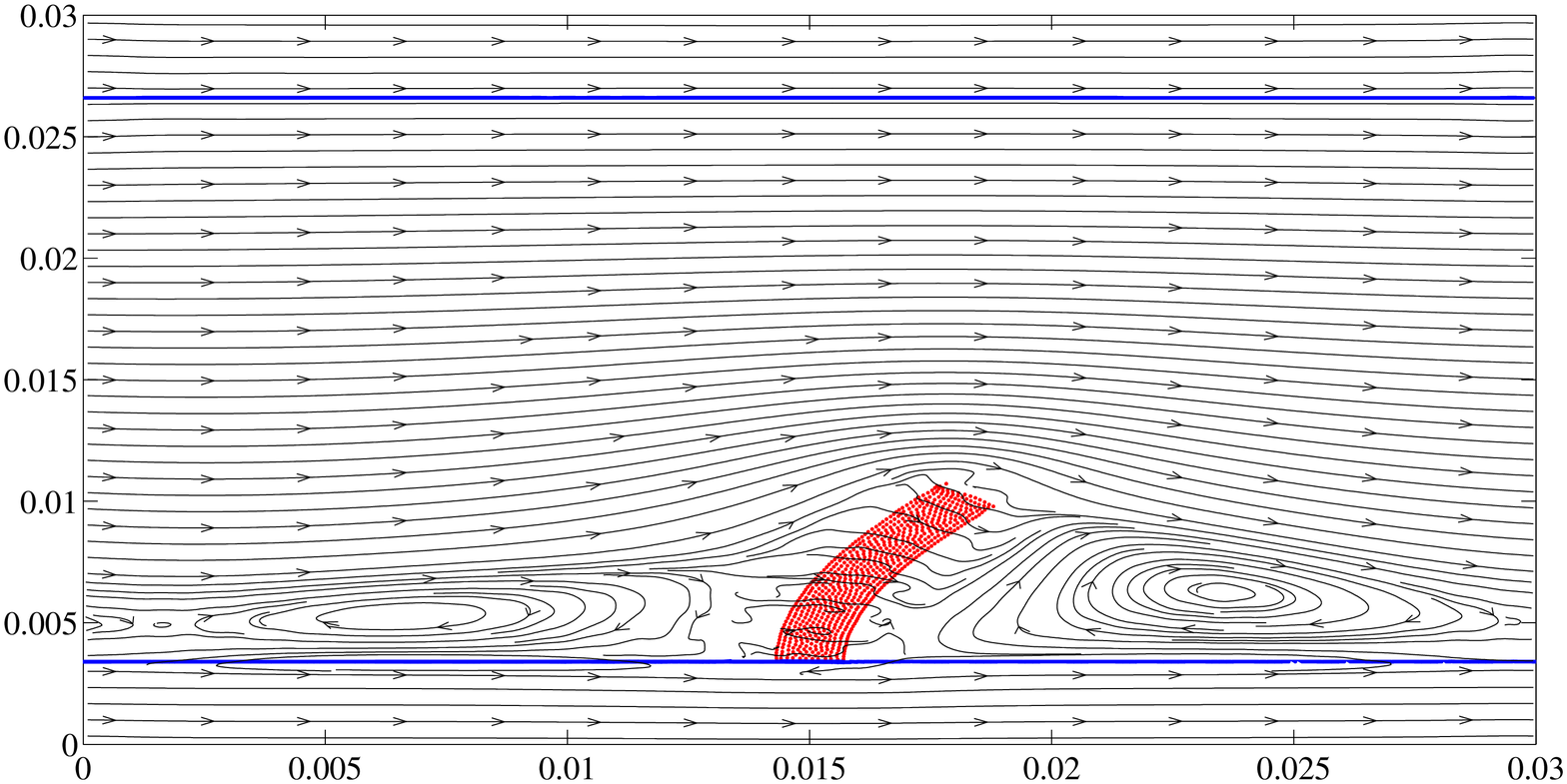}
    &
    \includegraphics[width=0.5\textwidth]{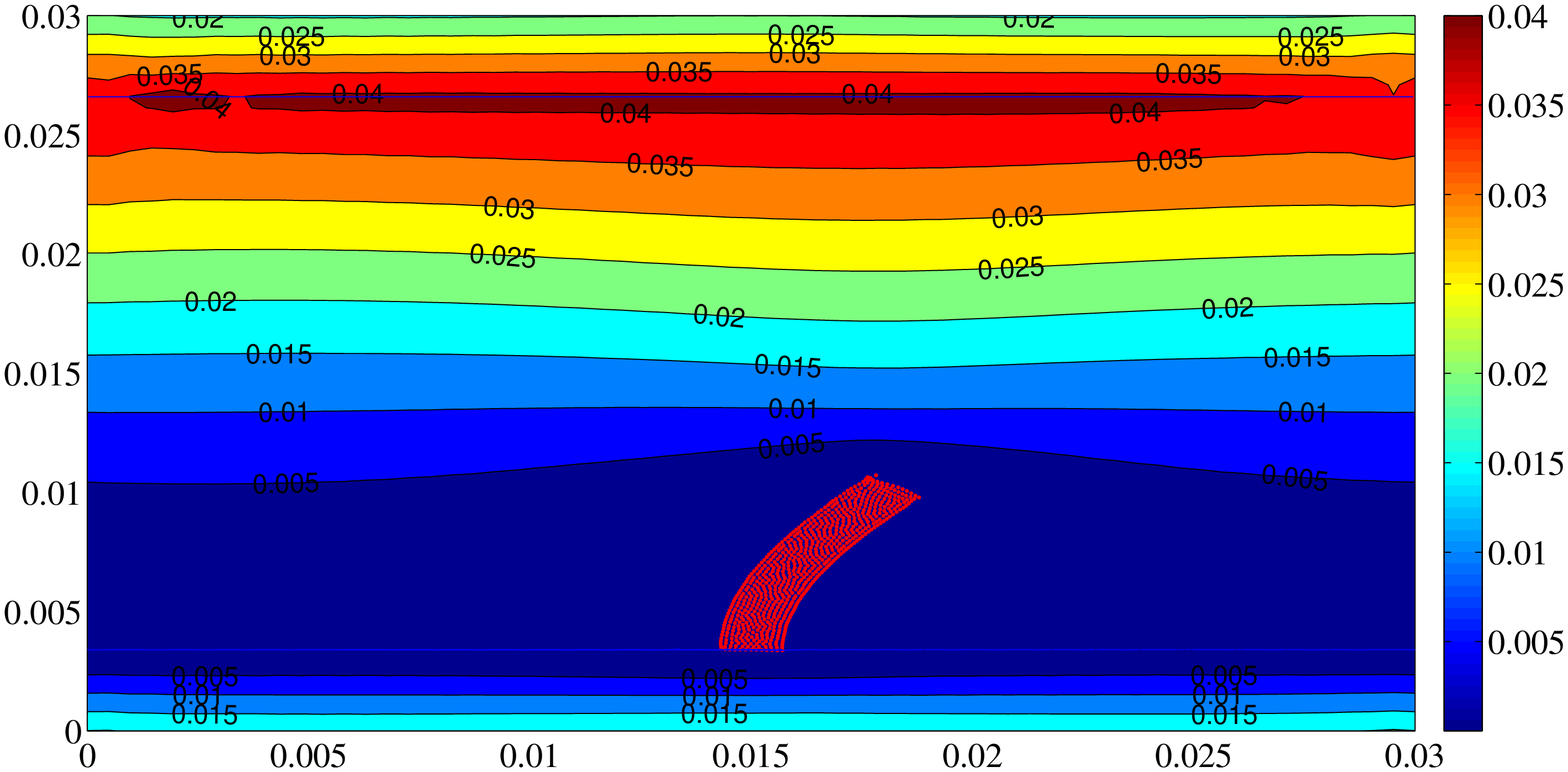}
  \end{tabular}
  \caption{Deflected beam for several values of $\utop$, showing
    streamlines (left) and contours of velocity magnitude (right).
    Other parameter values are $\Hbeam=0.0077\;\units{cm}$,
    $\sigma_b=560\;\units{g/cm\,s^2}$.}
  \label{fig:varyutop-stream}
\end{figure}

In our final sensitivity study, we vary the hydrodynamic force acting on
the beam by changing the top wall velocity $\utop$.  In contrast with
the previous cases, we consider much stronger shear flows corresponding
to $\utop$ values taken from the interval $[0.003, 0.6]$, which at the
upper end generates beam deformations that are well outside the linear
regime.  Figure~\ref{fig:varyutop-stream} depicts the deformed beam and
corresponding streamlines for values of $\utop=0.003,0.01,0.02,0.04$,
computed over long enough times  ($ t \approx 25 $s) that the beam has essentially reached a
steady state configuration.

If we consider the value of the last streamline lying above the beam tip (from Figure \ref{fig:varyutop-stream}), which
  yields the following values of $\|u\|$:
  \begin{center}
    \begin{tabular}{r|r|r}
      $\|u\|$ & $\utop$ & $\|u\|/\utop$ \\\hline
      0.0005  & 0.003   & 0.17\\
      0.001   &  0.01   & 0.1\\
      0.002   &  0.02   & 0.1\\
      0.005   &  0.04   & 0.125
    \end{tabular}
  \end{center}
  Consider the ratio of $\|u\|/\utop$ which is approximately 0.1 for all
  cases, suggesting a clear separation into a boundary layer region
  where the velocity drops roughly below 10\%\ of the top wall speed.

In each case, the flow separates roughly into two regions: a slow inner
flow near the wall that ``stalls'' ahead of and behind the beam; and a
much faster outer shear flow that bypasses the beam.  As the outer shear
flow increases in strength for higher $\utop$, the flow is characterized
by recirculating eddies ahead of and behind the beam.  We remark also
that streamlines pass through the beam and the wall, which at first
seems counter-intuitive if the immersed boundaries (and the adjacent
fluid) are supposed to be stationary at equilibrium.  However, it is
important to keep in mind that the IB points making up both the wall and
beam experience very small-amplitude oscillations even when the flow is
near steady state, and so there are still small non-zero velocities in
the neighbourhood of the immersed boundaries.

\leavethisout{
  \begin{note}[Choice of $\Hbeam$]
    Again, why did you choose $\Hbeam=0.0092$ and not some ``base''
    value from one of the previous two sections?  This scattershot
    approach to choosing parameters makes it seem like we haven't really
    put much thought into what values we're using.  Maybe you can put
    some thought into a ``testing plan'' that will use as many existing
    results as possible and so minimize the number of new simulations
    you need to run.  Keep in mind also that for the smoothed/porous
    beams in later sections, we will want to use the same parameters as
    for the rectangular/solid cases.
  \end{note}
}

\begin{figure}[tbhp]
  \centering
  \includegraphics[width=0.75\textwidth]{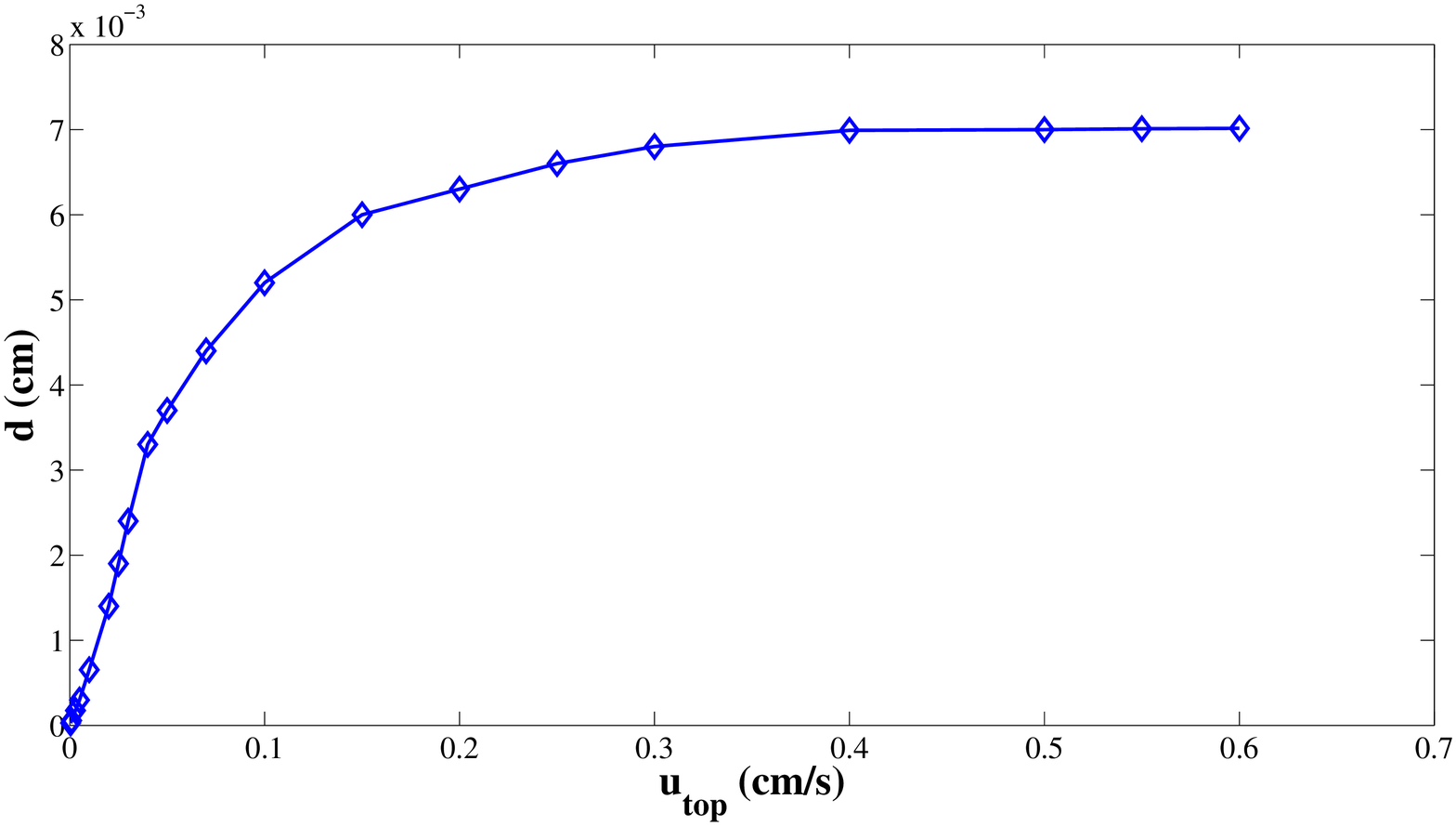} \\
  \caption{Maximum tip deflection as a function of $\utop$ for the solid
    beam, with parameters $\Hbeam=0.0077\;\units{cm}$ and
    $\sigma_b=560\;\units{g/cm\,s^2}$.}
  \label{fig:varyutop}
\end{figure}

The results of these simulations, as well as for a number of other flow
speeds, are summarized in Figure~\ref{fig:varyutop} on a plot of tip
deflection versus top wall velocity, which clearly shows that the beam
deformation can be divided into three regimes according to $\utop$:
\begin{itemize}
\item For $\utop\lessapprox 0.03$ or $\gamma \lessapprox 1.282$, the
  beam deflection is small enough that the linear beam theory applies
  and the load force varies roughly linearly with $\utop$.
\item For $\utop \gtrapprox 0.12$ or $\gamma \gtrapprox 5.128$, the beam
  has reached the maximum allowable deflection beyond which any
  additional increase in $\utop$ has little effect on the shape of the
  beam at steady-state.  This behaviour can be attributed to the fact
  that at larger values of the shear rate, the beam undergoes a
  significant vertical deflection as it bends downward and to the right,
  to the point where a large fraction of the beam is oriented parallel
  to the bulk flow.  This in turn reduces the net fluid load force
  acting on the beam to the point where increasing $\utop$ any further
  does not change the deflection (refer to
  Figure~\ref{fig:varyutop-stream}d).
\item For intermediate values of $0.03 \lessapprox \utop \lessapprox
  0.12$ or $1.282 \lessapprox \gamma \lessapprox 5.128$, the beam is in
  a transitional state between the low and high shear regimes described
  above.
\end{itemize}

\begin{figure}[tbhp]
  \centering
  \begin{minipage}{0.35\textwidth}
    \includegraphics[width=\textwidth]{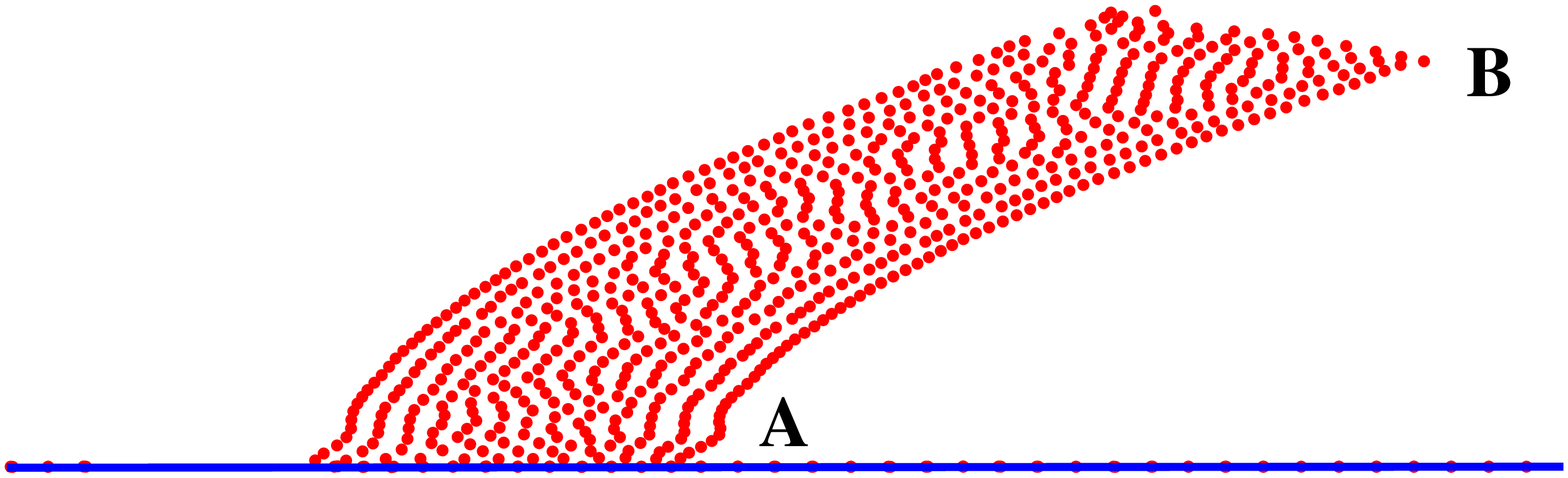}
  \end{minipage}
  \begin{minipage}{0.1\textwidth}
    \mbox{{\Huge \rotatebox{-18}{$\nearrow$}}\hspace*{0.15cm}\raisebox{0.5cm}{B}}\\[1cm]
    \mbox{{\Huge$\rightarrow$}\hspace*{0.35cm}\raisebox{0.1cm}{A}}
  \end{minipage}
  \begin{minipage}{0.45\textwidth}
    \fbox{\includegraphics[width=\textwidth]{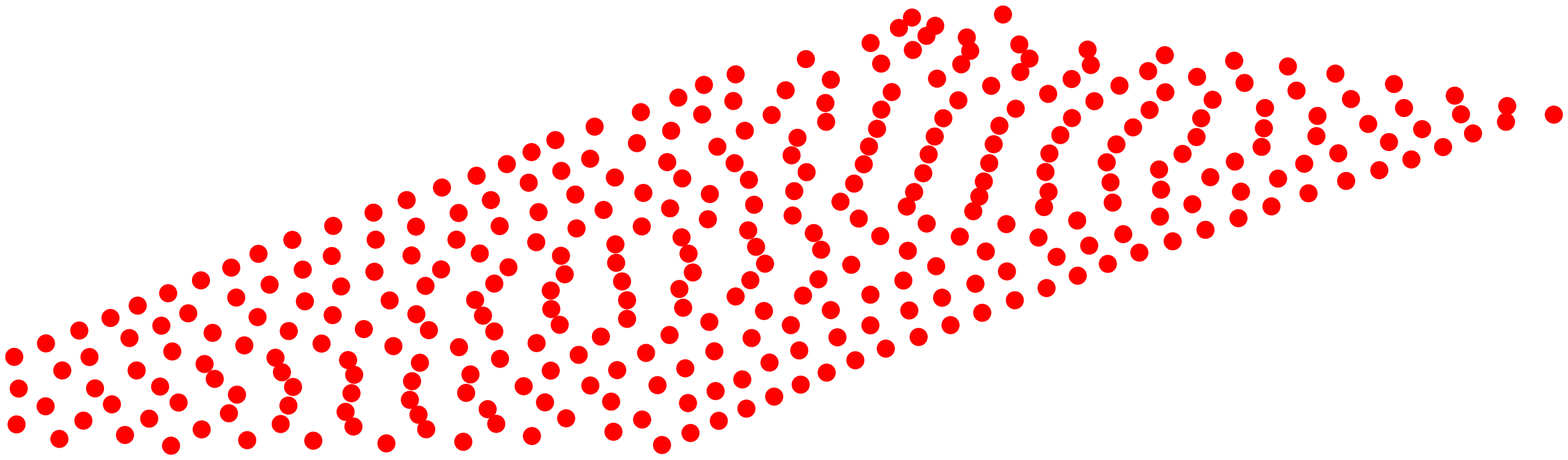}}\\[0.3cm]
    \fbox{\includegraphics[width=\textwidth]{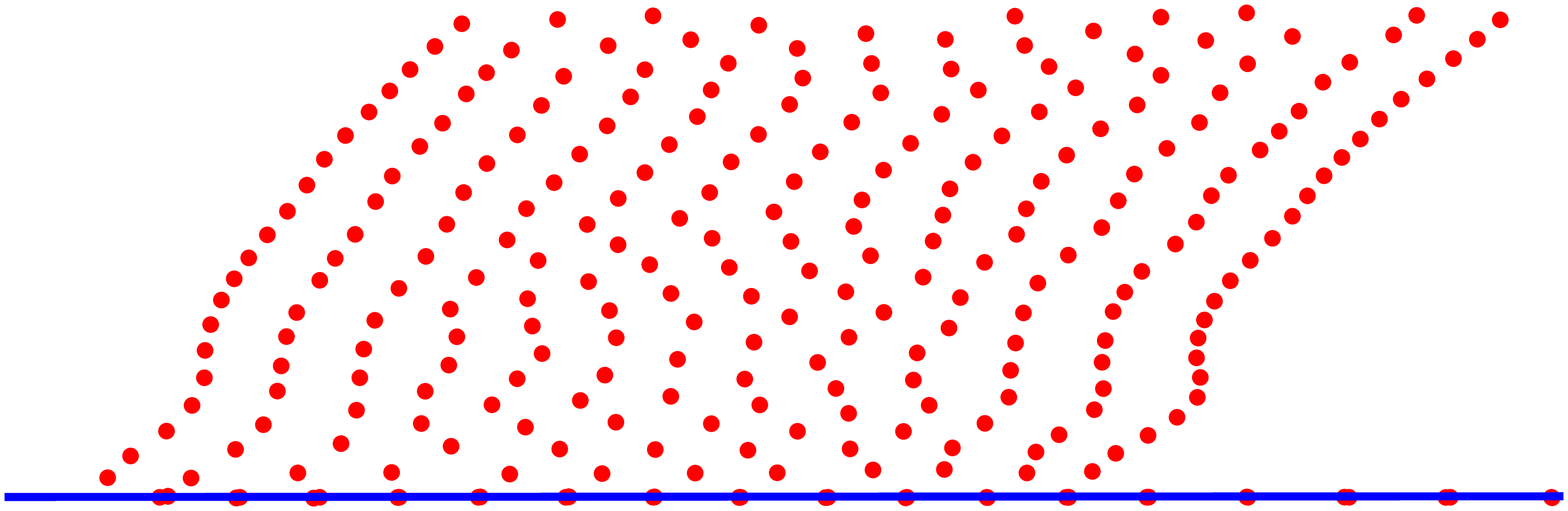}}
  \end{minipage}
  \caption{Zoomed view of the large deformations near the corners of the
    beam when $\utop=0.15$.}
  \label{fig:rect-zoomed}
\end{figure}

Another feature of the beam deformations that becomes evident at higher
$\utop$ is a large deviation from the original rectangular shape for the
IB points near either end of the beam, most notably at the corners.
This behaviour is emphasized in Figure~\ref{fig:rect-zoomed} which shows
two zoomed-in plots of the corner regions near the upper and lower ends
of the beam.  Near the wall (in the area denoted as region A) large
bending forces are generated that cause the beam to ``bow out'' on both
sides.  Furthermore, there are noticeable ``kinks'' at both bottom
corners involving the left- and right-most IB points immediately
adjacent to the lower wall.  At the free end of the beam (zoomed in as
region B) the left corner undergoes a significant distortion and the
corner point protrudes a significant distance out into the flow owing to
the very large shear forces experienced in that region.  Both of these
behaviours seem non-physical and would not be expected in an actual beam
with uniform elastic properties.  We attribute the anomalous
deformations near both ends of the beam to the existence of sharp
corners in the rectangular shape, and the purpose of the next section is to
investigate a modified beam shape that aims to eliminate
these anomalies.

\leavethisout{
  \begin{note}[Deformation plot]
    What parameter values did you use for the plot {\tt
      rect\_1} in Figure~\ref{fig:rect-zoomed}?
  \end{note}
}


\subsection{Smoothed cantilever beam}
\label{sec:smoothed-beam}

\begin{figure}[tbhp]
  \centering
  \begin{minipage}{0.45\textwidth}
    \includegraphics[width=1.1\textwidth]{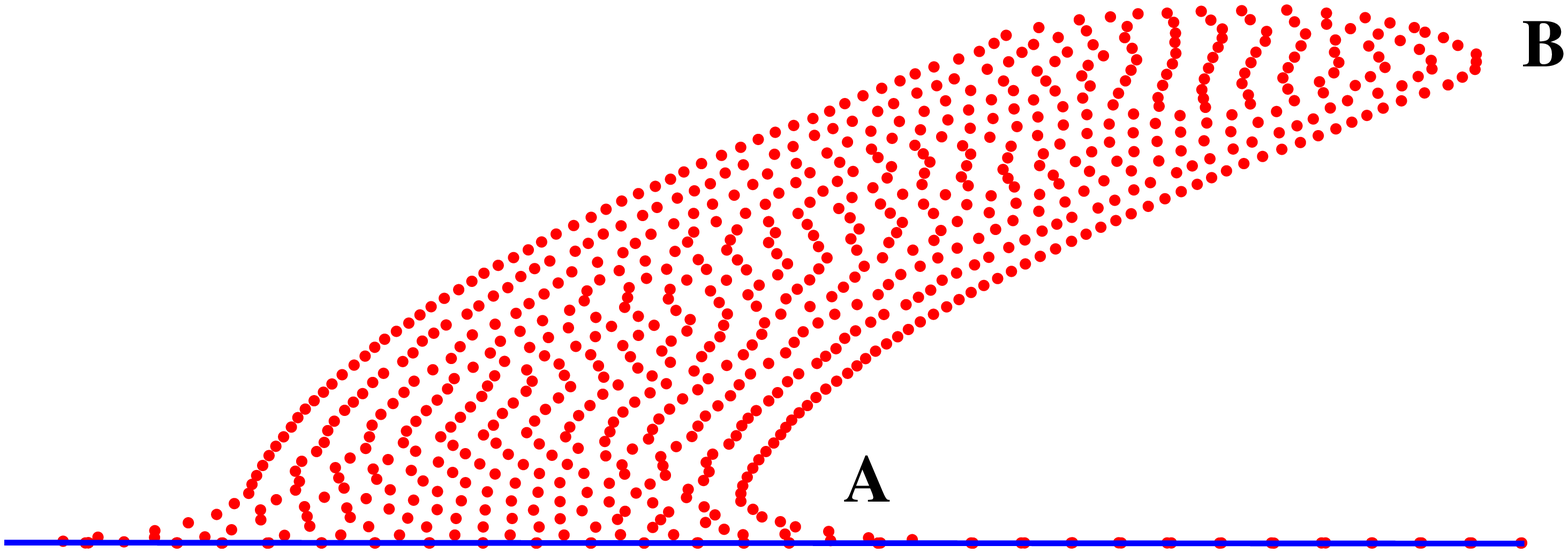}
  \end{minipage}
  \begin{minipage}{0.1\textwidth}
    \mbox{{\Huge \rotatebox{-18}{$\nearrow$}}\hspace*{0.15cm}\raisebox{0.5cm}{B}}\\[1cm]
    \mbox{{\Huge$\rightarrow$}\hspace*{0.35cm}\raisebox{0.1cm}{A}}
  \end{minipage}
  \begin{minipage}{0.4\textwidth}
    \fbox{\includegraphics[width=\textwidth]{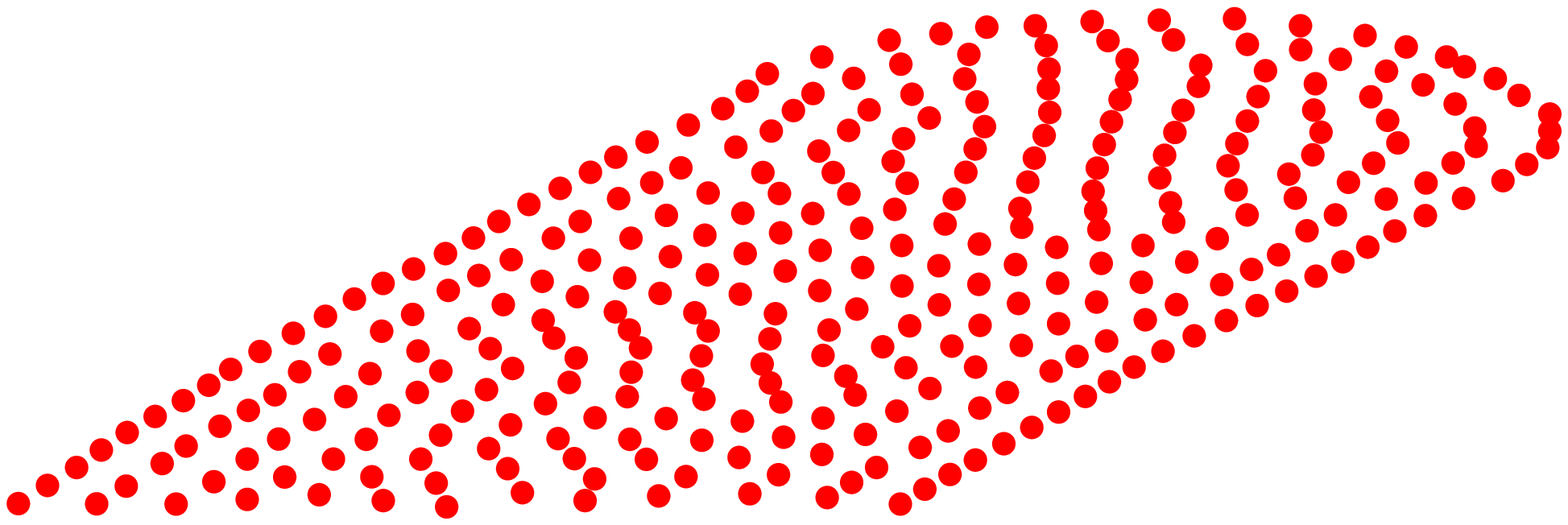}}\\[0.3cm]
    \fbox{\includegraphics[width=\textwidth]{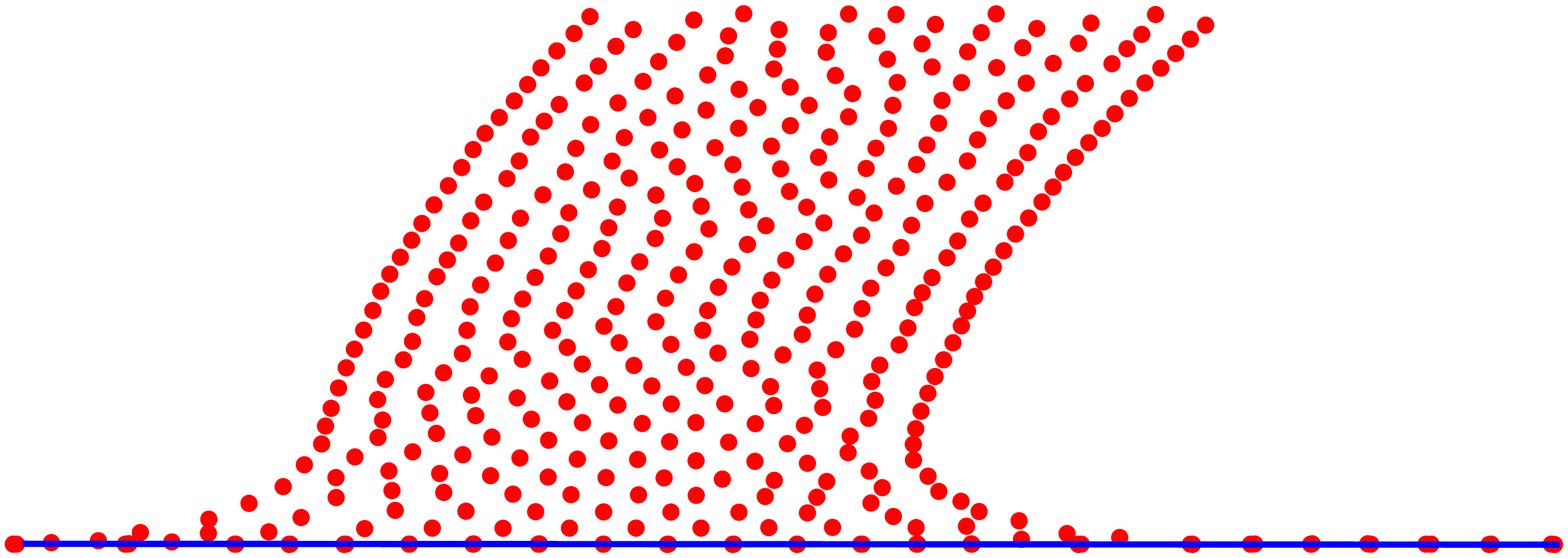}}
  \end{minipage}
  \caption{Zoomed view of the two ends of the smoothed beam when
    $\utop=0.15$.}
  \label{fig:smooth-zoomed}
\end{figure}

We next consider a modified beam shape that has the corners at both ends
smoothed out as shown in Figure~\ref{fig:dist_mesh}b using using fillets
or chamfers.  In particular, all four corners have been replaced using
circular arcs that have diameter equal to the beam width $\Wbeam$.  The
resulting shape is triangulated as before using {\tt distmesh2d}, and we
choose parameter values that are identical to the $\utop=0.15$
simulation from the previous section.  The resulting equilibrium beam
configuration is depicted in Figure~\ref{fig:smooth-zoomed}, which
exhibits a shape that is much more regular than in the rectangular case,
even for this high value of the fluid velocity.

\begin{figure}[tbhp]
  \centering
  \begin{tabular}{c}
    (a) \\
    \includegraphics[width=0.75\textwidth]{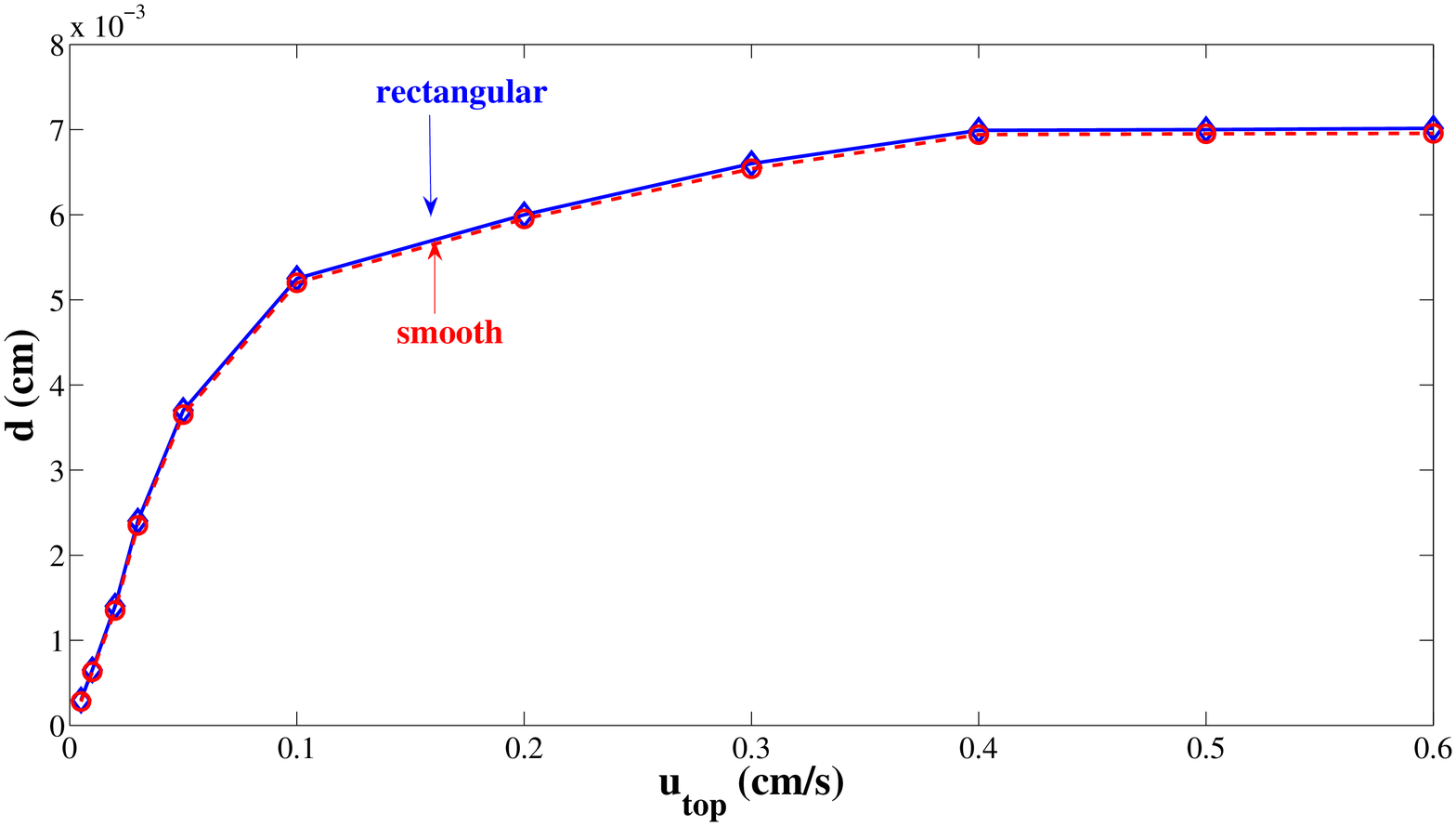} \\
    (b) \\
    \includegraphics[width=0.75\textwidth]{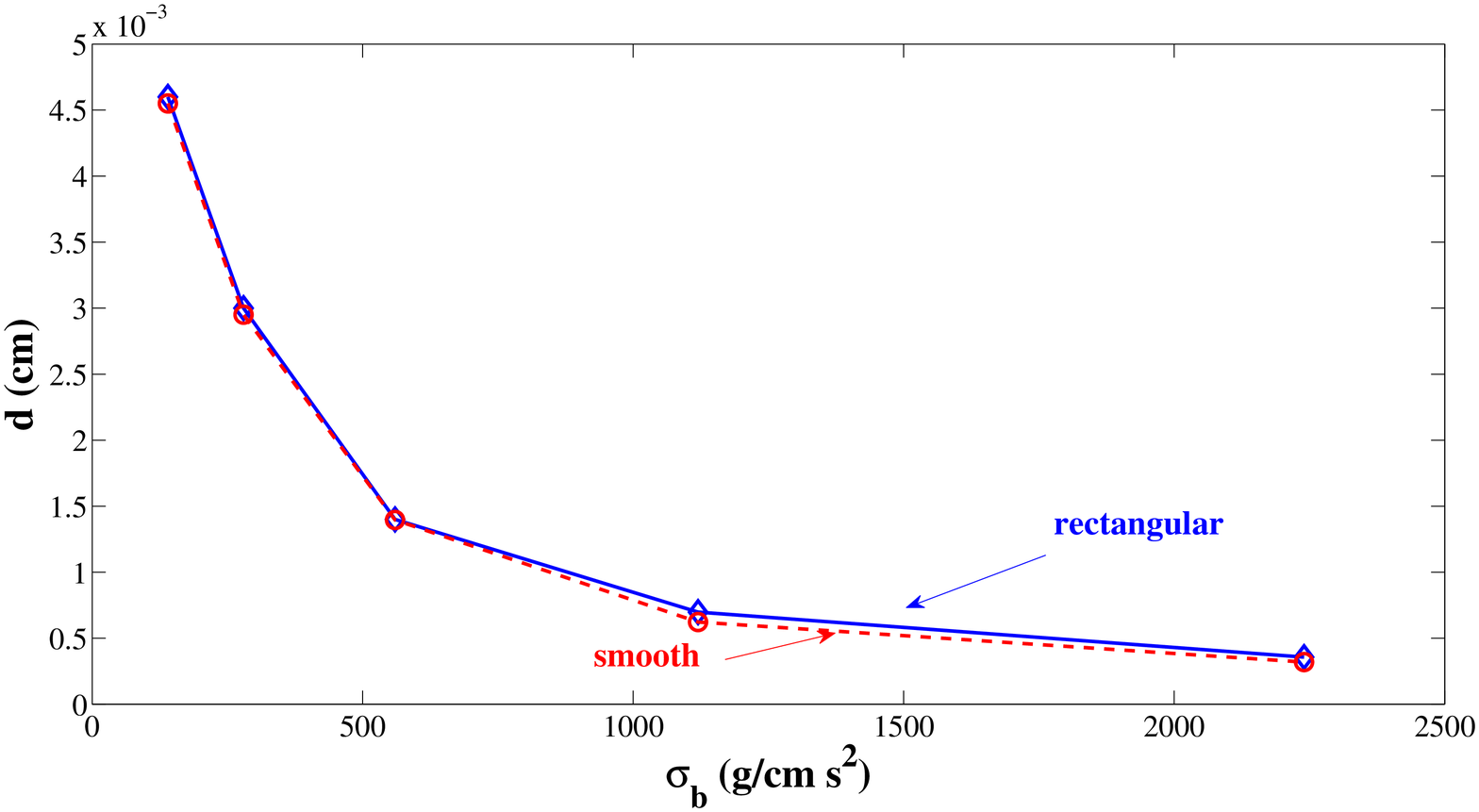}
  \end{tabular}
  \caption{Maximum tip deflection for the rectangular and smooth shapes,
    plotted as a function of (a) top wall velocity $\utop$ and (b) beam
    spring stiffness $\sigma_b$.}
  \label{fig:smooth-compare}
\end{figure}

Although the change in beam shape due to smoothing out of the corners
does influence the solution, these changes are very small as seen in the
two plots of tip deflection versus $\utop$ and $\sigma_b$ in
Figure~\ref{fig:smooth-compare}.  Indeed, the relative differences
between the two solutions is less than 1\%.  We will therefore employ
the smoothed beam shape in all simulations from this point onward.  And
with a view to applications, the rounded beam shape also has the
advantage that it is much more realistic in the context of biofilm
problems.



\section{Numerical simulations of a porous beam}
\label{sec:porous}

In this section, we consider a porous deformable structure, so that
fluid can flow through the structure in response to transmural pressure
gradients. Porous structures abound in biological systems, including
such examples as artery walls \cite{huang-tarbell-1997} and porous
biofilm layers \cite{vanloosdrecht-etal-2002,
  nguyen-morgenroth-eberl-2005, thullner-baveye-2007,
  davit-etal-2013}. Porous immersed boundaries also appear in
engineering applications such as filtration and separation processes.

Several efforts have been made to generalize the IB method to handle
porous structures. The first attempt to incorporate porosity within the
IB framework was a study of parachute dynamics by Kim and Peskin
\cite{kim-peskin-2006}, wherein the air vents at the apex on the chute
were dealt with by allowing the normal velocity of the canopy to differ
from that of the fluid by an amount proportional to the normal component
of the boundary force (which according to the jump conditions is also
proportional to the pressure jump). Stockie \cite{stockie-2009} built on
Kim and Peskin's approach by incorporating porosity directly using
Darcy's law. The IB method has also been used to study flow through
granular media at the pore scale by treating the grains making up the
medium as immersed boundaries \cite{dillon-fauci-2000}, although the
grains themselves are rigid and impermeable in these studies. Layton
\cite{layton-2006} generalized the closely related immersed interface
method by introducing a porous slip velocity in the direction normal to
the interface that is driven by differences in both transmural water
pressure and solute concentration.  In this thesis, we will extend the
ideas in \cite{kim-peskin-2006,stockie-2009} for a 1D porous membrane to
the case of a solid porous region in 2D. We will use this approach to
study the deformation of a porous rectangular cantilever beam, as well
as the same smoothed cantilever shape from the previous section.

\subsection{Extending the IB formulation to include porosity}

The system of equations \en{ns-mom}--\en{ib-velocity} is now generalized
to include the effect of porosity on the deformable structure. We follow
the approaches of Kim and Peskin \cite{kim-peskin-2006} and Stockie
\cite{stockie-2009} by incorporating a porous slip velocity in the fiber
evolution equation. The primary assumption made in \cite{stockie-2009}
and \cite{kim-peskin-2006} is that all the pores are directed normal to
the fiber. The slip velocity is given by $\vu_p = u_p\cdot\vn +
v_p\cdot\vt$ where $\vn$ and $\vt$ are the unit normal and tangential
vectors to the fiber. Due to the assumption just mentioned, the
tangential component $v_p = 0$. For a porous 2D region, it is not
realistic to continue with their assumption.

Indeed, we assume a more general porous structure with no restriction in the direction of the pores  and fluid can flow in any direction in the porous
region. The only way to determine the direction of the fluid velocity is
to look at the direction of the pressure gradient, $\nabla p$, that
drives the porous flow. In our approach, we also incorporate porous
effects via a porous slip velocity $\vu_p$; however here $\vu_p$ is
related to the pressure gradient (and not the pressure jump) via Darcy's
law:
\begin{gather}
  \vu_p = -\frac{K}{\mu} \nabla p,
\end{gather}
where $K$ represents the structure permeability [$cm^2$].

The porous beam obeys the same governing equations except that the
fiber evolution equation \en{ib-velocity} is replaced with
\begin{gather}
  \frac{\partial\vX}{\partial t} = - \vu_p + \int_\Omega
  \vu(\vx,t)~\delta(\vx-\vX(s,t))d\vx.
\end{gather}
The negative sign indicates that when the fluid moves through the porous
region outward with slip velocity $\vu_p$, the beam in turn moves in
the opposite direction, $-\vu_p$.

\begin{figure}[tbhp]
  \centering
  \begin{tabular}{m{0.18\textwidth}m{0.7\textwidth}}
    (a) $K=10^{-10}$ & \includegraphics[width=0.7\columnwidth]{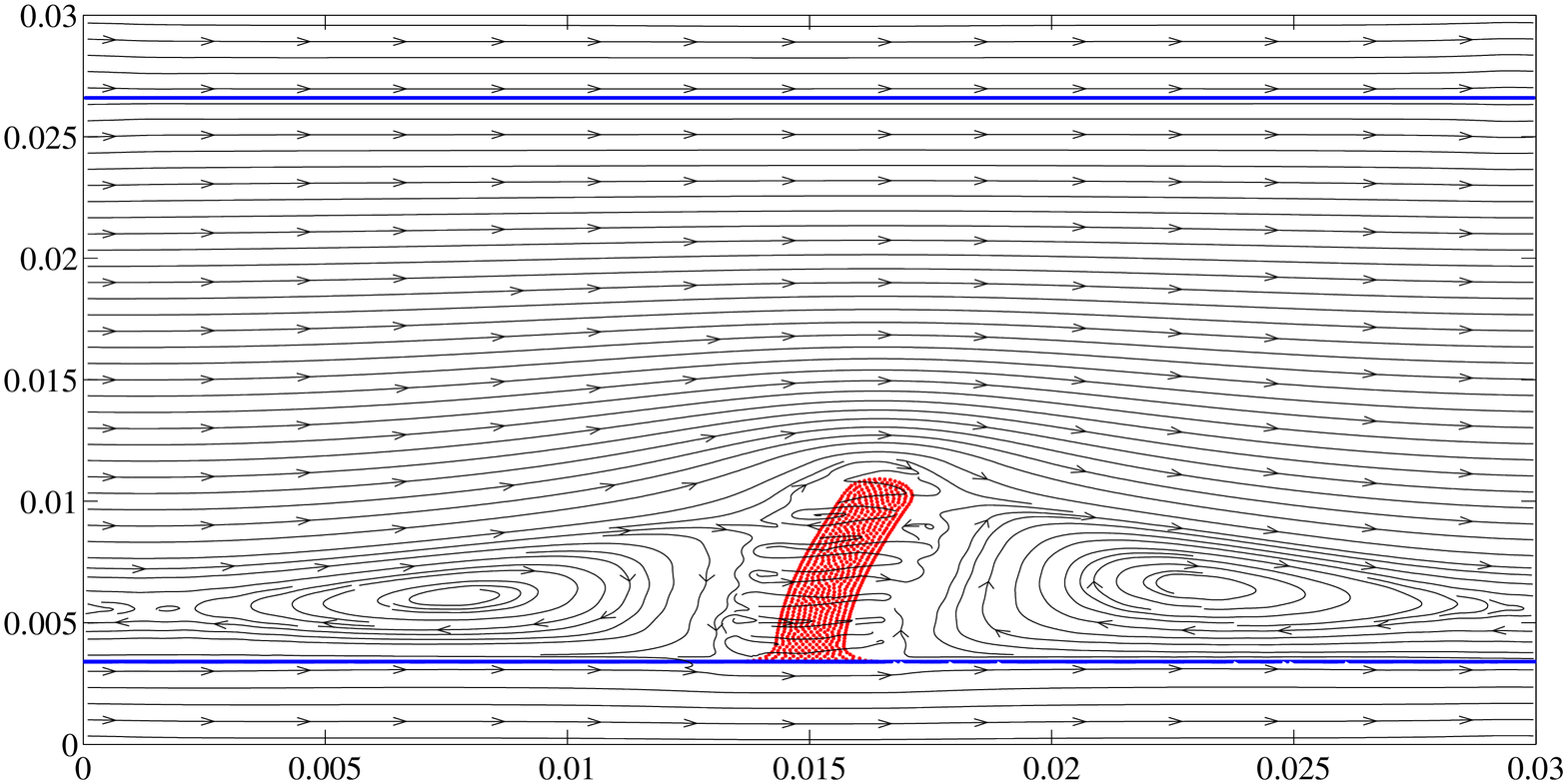}\\
    (b) $K=10^{-8}$  & \includegraphics[width=0.7\columnwidth]{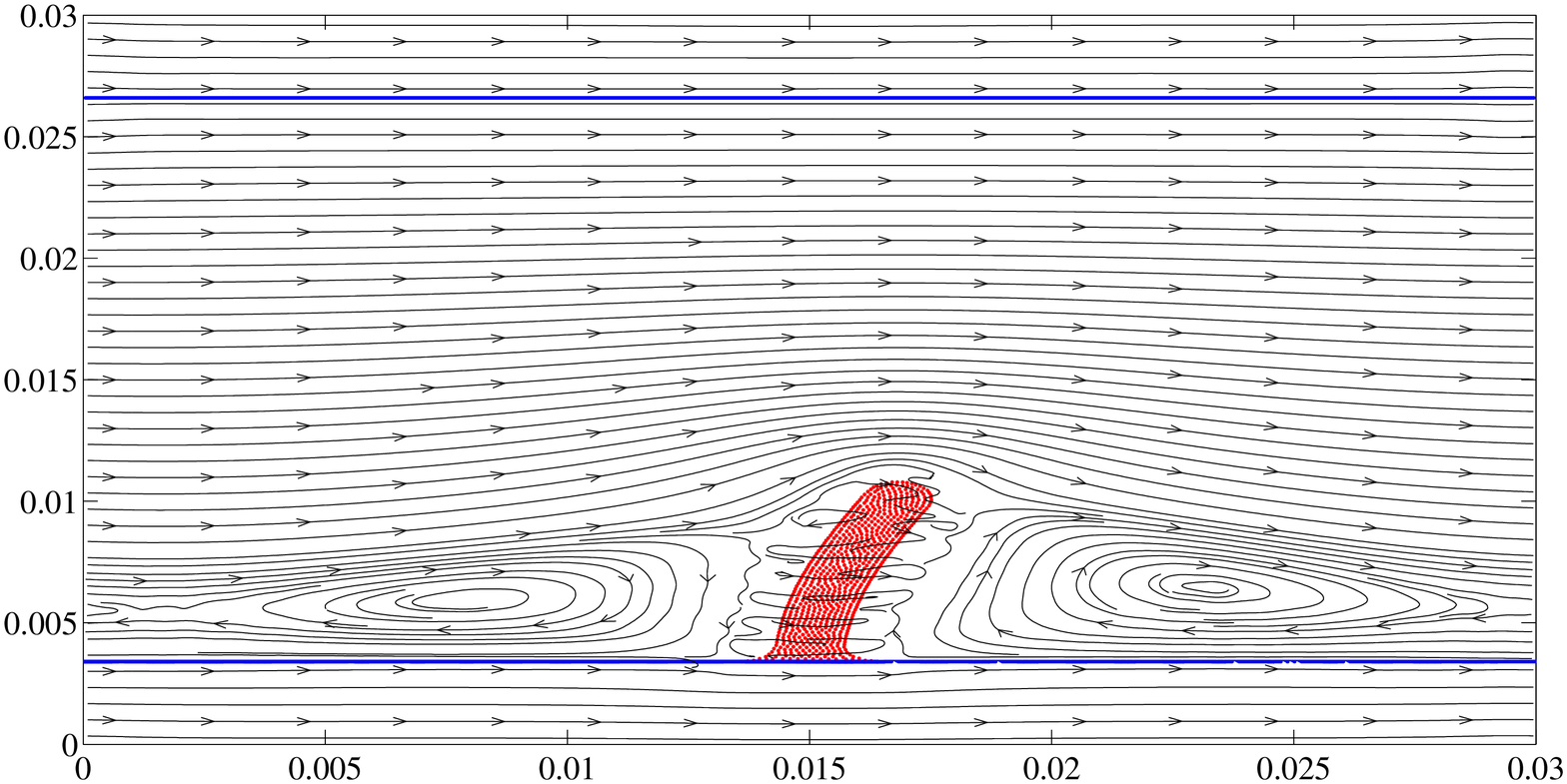}\\
    (c) $K=10^{-6}$  & \includegraphics[width=0.7\columnwidth]{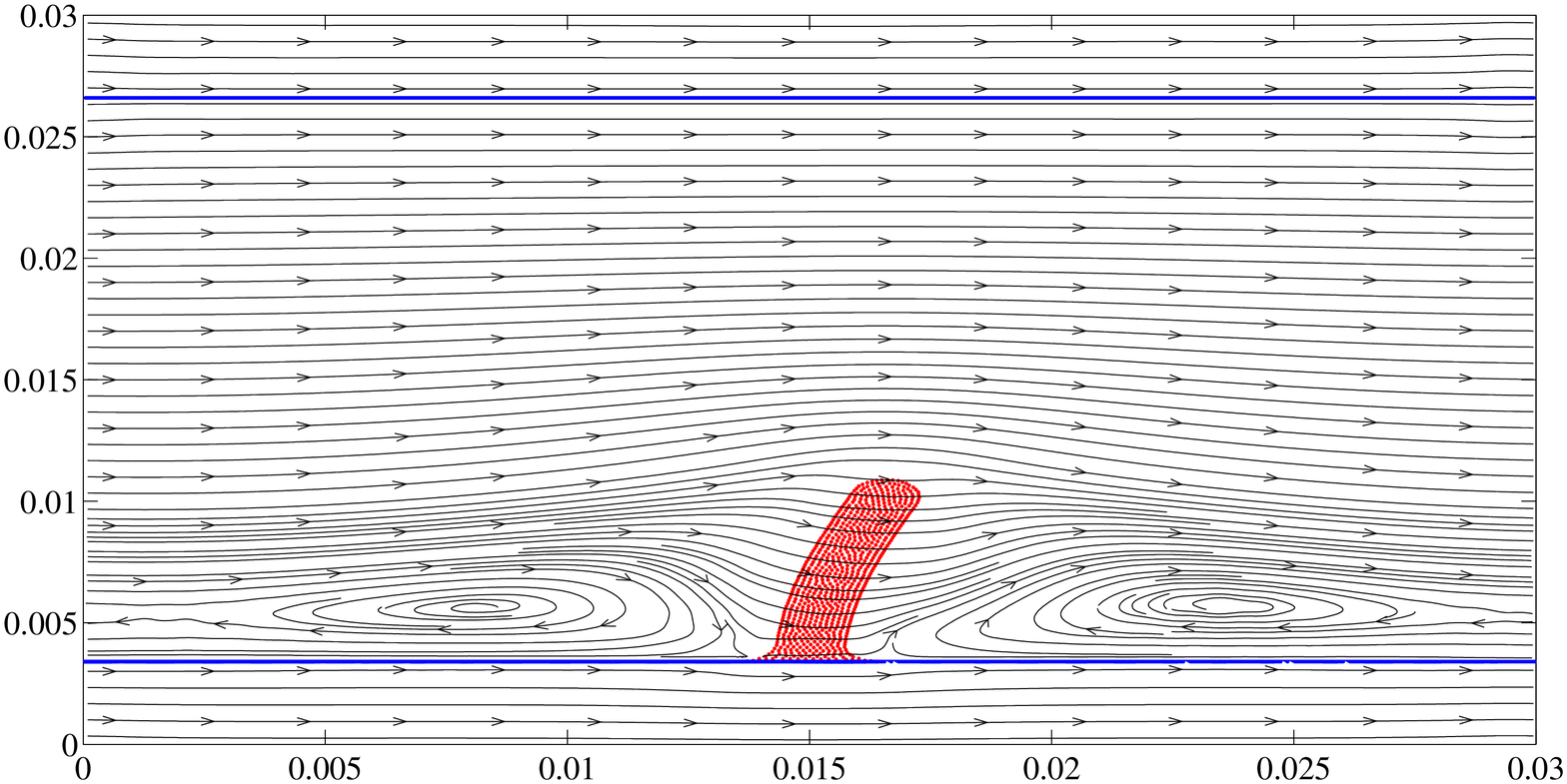}\\
    (d) $K=5\times 10^{-5}$ & \includegraphics[width=0.7\columnwidth]{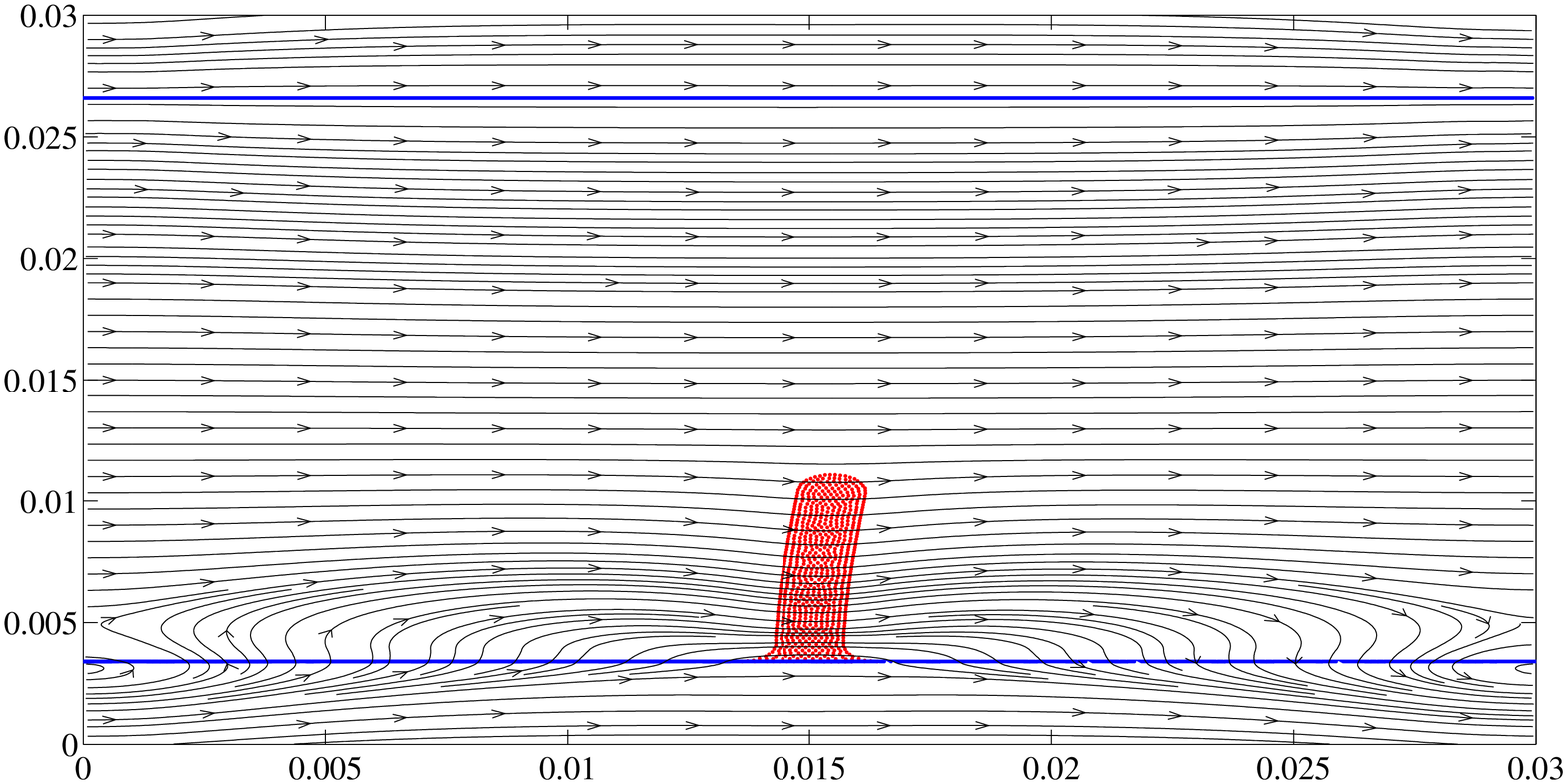}
  \end{tabular}
  \caption{Streamline plots at the point of maximum deflection for a
    smoothed porous beam with different permeability values.}
  \label{fig:streamline_canti}
\end{figure}

\begin{figure}[tbhp]
  \centering
  \includegraphics[width=0.75\textwidth]{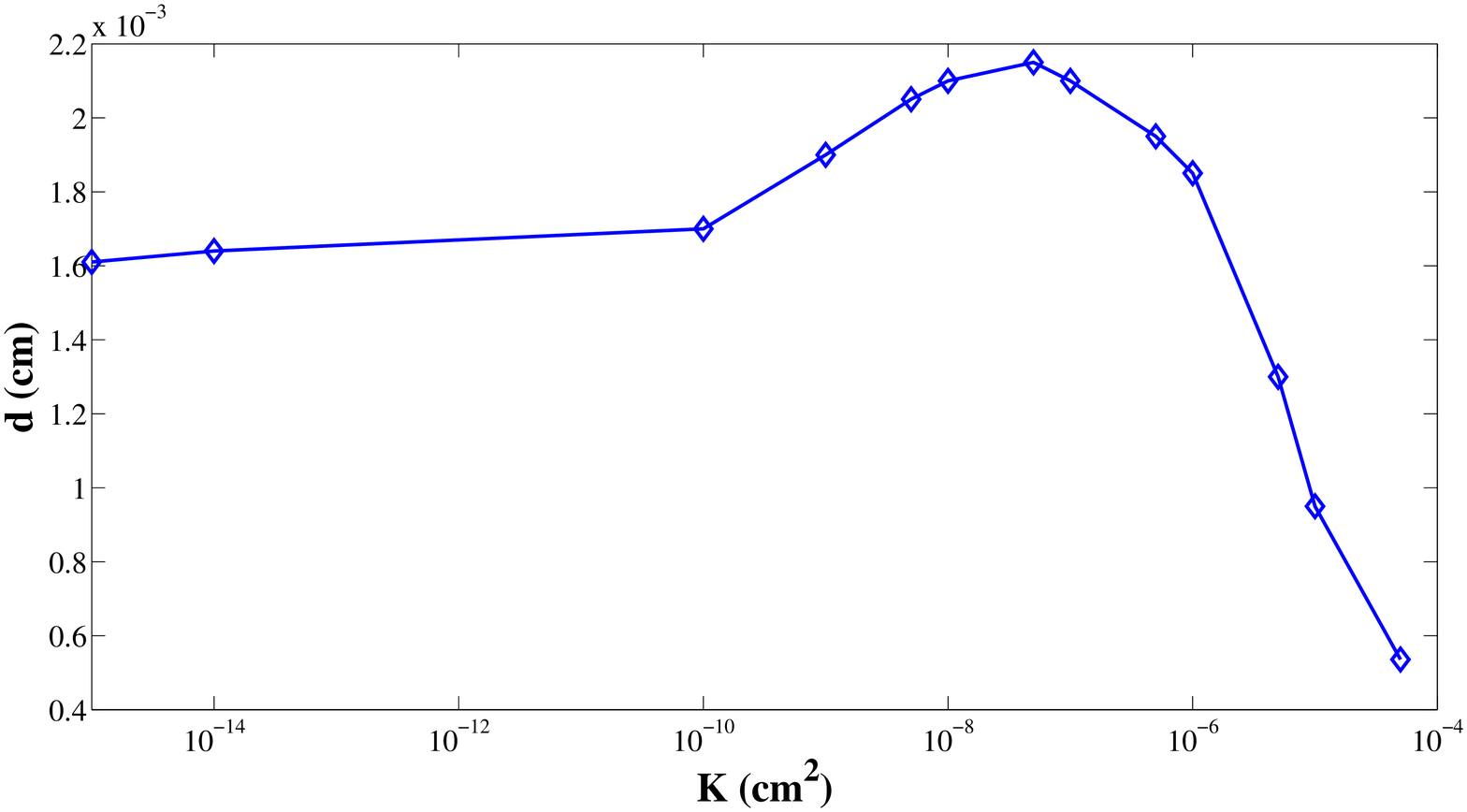}
  \caption{Final deflection ($\deflect$) versus permeability ($K$) for
    the smoothed porous beam.}
  \label{fig:cant_porosity}
\end{figure}

In our simulations, we have fixed values of $\Hbeam=0.0077\;\units{cm}$,
$\sigma_b=560\;\units{g/cm\,s^2}$ and $\utop=0.02\;\units{cm/s}$, which
corresponds to a shear rate of $0.0855\;\units{s^{-1}}$. Permeability $K$
lies in the range $[10^{-10}, 10^{-4}]\;\units{cm^2}$ and
Figure~\ref{fig:streamline_canti} shows the streamline plots for
different values of $K$. It is evident from the figure that for high
value of $K$, the fluid passes through the beam without any obstruction
but that is not the case for smaller values of $K$. When $K$ is taken as
small as $K=10^{-9}$, the flow is almost identical to that for the beam
in the non-porous case ($K=0$).  Figure \ref{fig:cant_porosity} shows how
the maximum deflection experienced by the cantilever varies with
permeability, from which we observe the following:
\begin{itemize}
\item For $K\lessapprox 10^{-8}$: the beam is essentially solid and the
  deflection is the same as in the non-porous case.
\item For $10^{-8}\lessapprox K\lessapprox10^{-6}$: there is a slight
  increase in maximum deflection. We ascribe this to an increase in
  porous flow through the beam that acts to enhance the horizontal
  component of velocity in the region near the beam which in turn
  increases the shearing force acting on the beam.
\item For $K\gtrapprox 10^{-6}$: the porous slip velocity is so large
  that fluid can pass through the solid structure more freely, hence
  decreasing the forces acting to deform the beam.
\end{itemize}

\section{Conclusions}
\label{sec:conclusions}

The main aim of this study was to derive a mathematical model and the
corresponding computational scheme to study the deflection of a
two-dimensional deformable elastic cantilever beam in response to a
surrounding viscous, incompressible fluid flow.  The choice of
cantilever was motivated by our efforts to model biofilm structures in
the near future. Our study included careful validation of the IB
approach so that in the future the model can be easily generalized to
irregular and highly deformable 3D structures with porosity, detachment
etc. We investigated how variations in physical and numerical parameters
change the effective material properties of the elastic beam and also
made a qualitative comparison of the results with linear beam theory. We
also paid attention to ``corner effects'' (irregularities in beam shape
near free end and near wall connection points) and showed how this can
be remedied by smoothing out the corners with a filled or round (a
standard technique in solid modelling to reduce stress, and in
aerodynamics to reduce interference drag). Finally we extended the
previous work done on porous IB membranes, by introducing porosity into
our deformable elastic beam. In our study we have not imposed any
restrictions on the orientation of the pores. The results obtained are
consistent with physical intuition. For example, physical intuition
suggests that for higher values of permeability, the fluid should pass
through the beam without obstruction as compared to smaller values of
permeability. The streamline plots obtained in this study also suggests
the same.  Along the same lines of argument we can say that as
permeability of the beam increases, the deflection of the beam should
decrease as the force acting to deform the beam decreases and this
argument is in good agreement with our numerical results.

In future, we aim to investigate the relationship $EI \propto \sigma_b$
by studying two analytical approximations. Our first approach will be by
finding a polynomial fit to ``typical'' computed load, and then deriving
the corresponding Euler-Bernoulli beam solution. A second method will
apply a simpler spring network and attempt to derive the equilibrium
configuration analytically.  We will also aim for irregular, deformable
shapes corresponding to biofilm columns and streamers. Last but not
least, there is a limitation in assuming a 2D geometry and so one of our
future aims will be to extend our IB model to 3D
cantilevers in the shape of deformed cylinders or parallelipipeds.

\bibliography{Beam_IB}

%







\end{document}